\documentclass[final,5p,times,twocolumn]{elsarticle}

\pdfoutput=1 

\usepackage{subcaption}
\newcommand{\RNum}[1]{\uppercase\expandafter{\romannumeral #1\relax}}
\usepackage[hidelinks]{hyperref}
\usepackage{booktabs}
\usepackage{tabularx}
\usepackage{adjustbox}
\usepackage{multirow}
\usepackage{xcolor}

\usepackage{amssymb}
\usepackage{amsmath}
\usepackage{amsthm}
\usepackage{mathrsfs}  

\newtheorem{remark}{Remark}

\usepackage{subfiles} 

\journal{journal of non-Newtonian fluid mechanics}

\begin{document}

\begin{frontmatter}

\title{Uncertainty quantification for the squeeze flow of generalized Newtonian fluids}

\affiliation[inst1]{organization={Eindhoven University of Technology, Department of Mechanical Engineering},
            addressline={Groene Loper 5},
            city={Eindhoven},
            postcode={5600 MB}, 
            country={The Netherlands}}

\author[inst1]{Aricia Rinkens\corref{cor}}
\ead{a.rinkens@tue.nl}
\author[inst1]{Clemens V. Verhoosel}
\ead{c.v.verhoosel@tue.nl}
\author[inst1]{Nick O. Jaensson}
\ead{n.o.jaensson@tue.nl}

\cortext[cor]{Corresponding author: A.\,Rinkens ({a.rinkens@tue.nl})}

\date{July 2023}

\begin{abstract}
The calibration of rheological parameters in the modeling of complex flows of non-Newtonian fluids can be a daunting task. In this paper we demonstrate how the framework of Uncertainty Quantification (UQ) can be used to improve the predictive capabilities of rheological models in such flow scenarios. For this demonstration, we consider the squeeze flow of generalized Newtonian fluids. To systematically study uncertainties, we have developed a tailored squeeze flow setup, which we have used to perform experiments with glycerol and PVP solution. To mimic these experiments, we have developed a three-region truncated power law model, which can be evaluated semi-analytically. This fast-to-evaluate model enables us to consider uncertainty propagation and Bayesian inference using (Markov chain) Monte Carlo techniques. We demonstrate that with prior information obtained from dedicated experiments -- most importantly rheological measurements -- the truncated power law model can adequately predict the experimental results. We observe that when the squeeze flow experiments are incorporated in the analysis in the case of Bayesian inference, this leads to an update of the prior information on the rheological parameters, giving evidence of the need for recalibration in the considered complex flow scenario. In the process of Bayesian inference we also obtain information on quantities of interest that are not directly observable in the experimental data, such as the spatial distribution of the three flow regimes. In this way, besides improving the predictive capabilities of the model, the uncertainty quantification framework enhances the insight into complex flow scenarios. 
\end{abstract}


\end{frontmatter}

\section{Introduction}
\label{sec:intro}

Non-Newtonian fluids are encountered in many industrial applications such as food processing, additive manufacturing and coating. In these applications, different manufacturing
techniques are used, \emph{e.g.}, film blowing, injection moulding and extrusion. To understand and optimize these processes, models that accurately predict the involved complex flows, \emph{i.e.}, flows that consist of mixed, time-dependent and spatially varying deformation modes, are of the utmost importance \cite{Owens2002ComputationalRheology}.

The traditional modeling approach for complex flows assumes that model parameters are ``deterministic'', \emph{i.e.}, they are known exactly. Typically, these parameters are obtained by \mbox{(non-)linear} rheological measurements in well-defined flows, such as simple shear. Based on the measurements and external factors, such as experts' knowledge, intuition and experience, a constitutive model is selected, which is calibrated using the rheological data \cite{Macosko1996RheologyPrinciples}. The model is subsequently used to predict complex flow behavior.

An accurate prediction of fluid flows with complex rheological behavior requires sophisticated constitutive models \cite{Owens2002ComputationalRheology}. When calibrated well, such models can replicate the measured behavior in simple flow (as opposed to complex flows defined above) with a high accuracy. However, calibrating rheological parameters using rheological measurements is a non-unique procedure \cite{Singh2019OnRepresentation, Singh2022OnCertainty}. In addition, since the deformation types and flow history can be significantly different in complex flow scenarios, using the calibrated model in such scenarios does not necessarily lead to accurate predictions. 
Moreover, if the flow is surrounded by uncertainties, \emph{e.g.}, from operating conditions and flow behavior, predictions using the traditional modeling approach may be complicated further.

Uncertainty quantification (UQ) provides an alternative modeling framework for making predictions in settings where it is hard to accurately determine model parameters. In this ``probabilistic'' approach, it is assumed that the model parameters are stochastic, acknowledging the limited knowledge we have about them. The parametric uncertainties are determined from measurements, intuition or experts' knowledge. 
In UQ, the main goal is to quantify the uncertainty of the model predictions, as opposed to obtaining the most accurate predictions possible. Whether or not the uncertainty is acceptable depends on the considered problem.

There are several important aspects to UQ, which include mapping out experimental uncertainties, 
sensitivity analysis, uncertainty propagation and inference. Mapping out the uncertainties indicates whether
model parameters should be considered stochastic or deterministic.
Sensitivity analysis enables us to determine which parameters have the largest impact on the model prediction  \cite{Smith2013UncertaintyApplications}. In uncertainty propagation, we propagate the parametric uncertainties through the physical model to obtain a probabilistic model prediction. In inference, we incorporate experimental data from the complex flow to investigate the uncertainty in the model parameters with the goal of improving the probabilistic prediction, by updating the model parameters.

The concept of uncertainty quantification has been widely used in, \emph{e.g.}, nuclear reactor models \cite{Avramova2010VerificationAnalysis}, weather models \cite{Moosavi2021MachineModels}, biological models \cite{Wentworth2018BayesianAlgorithms} and solid mechanics \cite{Rappel2020AMechanics}. An extensive review for Bayesian inference in physics is given by Von Toussaint \cite{VonToussaint2011BayesianPhysics}. In the scope of non-Newtonian fluids mechanics and rheology, uncertainty quantification is relatively unexplored.
In complex flows, uncertainty propagation has been used to evaluate the uncertainty of a quantity of interest. In a study by Pereira \emph{et al.} \cite{Pereira2013OnModels} uncertainty propagation has been used, where the uncertainty in blood viscosity is propagated through the momentum balance to obtain quantities of interest such as the wall shear stress. Kim \emph{et al.} \cite{Kim2019UncertaintyFlows} have also used uncertainty propagation to predict the resistance and velocity of a settling sphere in a Carbopol and PVP solution, where the rheological parameters were obtained using Bayesian inference. Kumar \emph{et al.} \cite{Kumar2021Physics-guidedResources} have used deep neural networks (DNN) to characterize a non-Newtonian fluid and have applied uncertainty propagation to investigate the dominant parameters that affect the simulation. In a study performed by Sen \emph{et al.} \cite{Garcia2022VerificationFlows} uncertainty propagation has been applied to Newtonian and non-Newtonian fluid flows on the microscale level.

Uncertainty quantification in the form of Bayesian inference has been mainly applied to rheological studies. In a study performed by Freund and Ewoldt \cite{Freund2015QuantitativeInference}, Bayesian inference has been used for UQ and model selection in linear rheology. Ran \emph{et al.} \cite{Ran2023UnderstandingInference} have applied Bayesian inference in the rheological characterization of a kaolinite clay suspension, which can be modeled using a micro-structural viscoelastic model. Bayesian inference has also been used in the prediction of linear viscoelastic models of branched polymers in a study performed by Shanbhag \cite{Shanbhag2010AnalyticalFormulation}.
UQ has also been used to estimate rheological parameters in related fields such as 
geophysical research \cite{Hilley2005BayesianTibet,Korenaga2008ARheology}.

Sensitivity analysis in a complex flow setting has been performed by Freund \emph{et al.} \cite{Freund2018FieldParameters}, who studied field sensitivities of flow predictions to rheological parameters for generalized Newtonian and thixotropic fluids. This work was later extended to viscoelastic fluids at a low Deborah number by Kim \cite{Kim2023Adjoint-basedNumber}.

To summarize, uncertainty quantification has been applied in the field of non-Newtonian fluid mechanics in the form of sensitivity analysis, uncertainty propagation and Bayesian inference. However, inference has been mainly limited to simple flows as found in rheology, for which fast-to-evaluate models are available. To the best of our knowledge, no research has yet been done on applying Bayesian inference in a complex flow of non-Newtonian fluids.

In this work, we use the methodology of uncertainty quantification in the context of a complex flow case with a generalized Newtonian fluid. The study of this setting introduces two requirements. 
First, we need full control of our experiments, allowing for detailed assessment of the parametric uncertainties and model calibration. Therefore, we consider a squeeze flow for which we have developed a tailored experimental setup. A squeeze flow consists of extensional and shear deformations, making it a complex flow which is still computationally tractable. 
Second, the computation time for our physical model should be kept to a minimum, because it has to be evaluated many times to allow for uncertainty propagation and inference. To this end, we develop a semi-analytical model that allows for describing Newtonian and shear thinning flow behavior.

This manuscript is outlined as follows. To set uncertainty quantification in the scope of the current work, in Section~\ref{sec:uq} we first introduce the key concepts and terminology for UQ. In Section~\ref{sec:setup} we then introduce the squeeze flow setup and the corresponding experiments. Section~\ref{sec:model} describes the Newtonian and shear thinning model to mimic the experiment. In Section~\ref{sec:uncertainties} we describe the quantification of the parametric uncertainties. We then discuss the comparison between the squeeze flow model and experiments using uncertainty propagation and Bayesian inference in Section~\ref{sec:results}. Finally, we present the conclusions and recommendations in Section~\ref{sec:conclusion}.

\section{Uncertainty quantification}
\label{sec:uq}

In this section we outline the fundamental ideas of uncertainty quantification, introducing the terminology and notation used throughout this work. We refer the reader to, \emph{e.g.}, Oden \emph{et al.} \cite{Oden2017PredictiveUncertainty} for a more detailed exposition.

\subsection{Uncertainty modeling}
A fundamental goal of science is to make predictions of events in physical reality, or, more precisely, about quantities of interest related to such events. In modern science and engineering, the systems for which predictions are required are becoming more complex. The complexity of the systems gives rise to uncertainties in experimental observations used for model validation and for the calibration of model parameters, as well as uncertainties in the models themselves.

Uncertainty quantification (UQ) aims to make predictions of events in physical reality using models and taking into account the uncertainties that are inherent to the problem. A particular challenge in many engineering problems is that calibration of models using relatively simple experiments insufficiently reduces the uncertainties associated with the predictions for the real system of interest.

In UQ, quantities that are uncertain are described by random variables, which are characterized by probability distributions. To formalize this concept, we consider a distribution $\boldsymbol{g}$ of a vector of quantities that describe the ``true'' event, and experimental observations $\boldsymbol{y}=\{\boldsymbol{y}_1, ..., \boldsymbol{y}_n\}$ of this event,
where $n$ represents the number of observations. Denoting the realization of reality corresponding to the $i$-th observation $\boldsymbol{y}_i$ as $\boldsymbol{g}_i$ allows us to express the observation error $\boldsymbol{\varepsilon}_i$ (in an additive way) as 
\begin{equation}
    \boldsymbol{\varepsilon}_i = \boldsymbol{y}_i - \boldsymbol{g}_i.
    \label{eq:observationerror}
\end{equation}
We note that the truth, $\boldsymbol{g}_i$, is unknown in practice and that in general a model should be assumed for the observational error. Such a model is generally referred to as a noise model.

Similar to the definition of the observation error, we can also define a model error (or model bias) $\boldsymbol{\gamma}_i(\boldsymbol{\theta})$ as 
\begin{equation}    
    \boldsymbol{\gamma}_i(\boldsymbol{\theta}) = \boldsymbol{g}_i - \boldsymbol{d}_i(\boldsymbol{\theta}),
    \label{eq:modelbias}
\end{equation}
 where $\boldsymbol{d}_i$ is the model prediction corresponding to the parameter set $\boldsymbol{\theta}$. We define the parameter domain as $\boldsymbol{\Theta}=\Theta_1 \times \Theta_2 \times  ... \times \Theta_p$, where $\boldsymbol{\theta} \in \boldsymbol{\Theta}$ and $p$ is the number of parameters.

Combining the observational error \eqref{eq:observationerror} and the model bias \eqref{eq:modelbias} enables the elimination of the unknown truth from the error definition as 
\begin{equation}\label{eq:obserrmodelbias}
    \boldsymbol{y}_i - \boldsymbol{d}_i(\boldsymbol{\theta}) = \boldsymbol{\varepsilon}_i + \boldsymbol{\gamma}_i(\boldsymbol{\theta}).
\end{equation}
This expression conveys that the difference between the model predictions and the observations is a combination of two error contributions, \emph{viz.} the model errors and the observational errors. A fundamental aspect of uncertainty quantification is to postulate models of the error contributions, as these models play an essential role in the uncertainty in the problem. In practice, it is frequently decided to combine the two error contributions in equation~\eqref{eq:obserrmodelbias} in a single noise model, encompassing both error sources.

In this work we consider both experiments and models, which enables us to systematically study both error contributions. In particular, we study the relation between the simple calibration experiments, \emph{i.e.,} the rheological measurements, and the more complex application, \emph{i.e.,} a squeeze flow. Before considering the problem of interest in this work, we first discuss the methods used in the framework of UQ.

\subsection{Uncertainty propagation and inference}
The two main approaches of UQ are uncertainty propagation and inference. In uncertainty propagation, we quantify the uncertainty in model parameters, \emph{i.e.,} parametric uncertainties, $\boldsymbol{\theta}$, and propagate these through a physical model, $\boldsymbol{d}$, to obtain the uncertainty in the quantity of interest. In inference problems, the model parameters are updated based on observations. We evaluate the posterior probability density function, $\pi(\boldsymbol{\theta}|\boldsymbol{d})$, of the parameters based on both prior information of these parameters, $\pi_0(\boldsymbol{\theta})$, and on observations $\boldsymbol{y}$, using Bayes' rule    
    \begin{equation}\label{eq:bayesrule}
        \pi(\boldsymbol{\theta}|\boldsymbol{y}) = \frac{\pi(\boldsymbol{y}|\boldsymbol{\theta}) \pi_0(\boldsymbol{\theta})}{\pi_\Upsilon(\boldsymbol{y})}.
    \end{equation}
In this expression, the normalization constant, $\pi_\Upsilon(\boldsymbol{y})$, is referred to as the evidence. The posterior quantifies the probability of obtaining parameter values, $\boldsymbol{\theta}$, given the experimental data $\boldsymbol{y}$. The observations are encoded in the likelihood function, $L(\boldsymbol{\theta}|\boldsymbol{y}) \equiv \pi(\boldsymbol{y}|\boldsymbol{\theta})$.

By combining prior knowledge with experimental data, an improved model prediction is obtained compared to uncertainty propagation. The experimental data influences the posterior density through the likelihood term $L(\boldsymbol{\theta}|\boldsymbol{y})$, which quantifies the likelihood of observing the data $\boldsymbol{y}$ given the parameters $\boldsymbol{\theta}$. In the likelihood function, the measured data $\boldsymbol{y}$ is fixed and the parameters $\boldsymbol{\theta}$ are varied over the admissible domain. Because it is assumed that the data remains fixed and the parameter values can vary, the likelihood is not a probability distribution.
The model errors and observational errors, as discussed in relation to equation \eqref{eq:obserrmodelbias}, are incorporated in the likelihood function as random variables. A common way to do this is to assume the combined errors to be independent and identically distributed, following a normal distribution. We then obtain
\begin{equation}\label{eq:likelihood_general}
    L(\boldsymbol{\theta}| \boldsymbol{y}) = \frac{1}{(2 \pi \boldsymbol{\Sigma}_i)^{n/2}} \exp\left(- \frac{1}{2} \sum_{i=1}^n \left( \boldsymbol{y}_i - \boldsymbol{d}_i \right)^\text{T} \boldsymbol{\Sigma}_i^{-1} \left( \boldsymbol{y}_i - \boldsymbol{d}_i \right) \right),
\end{equation}
where $\boldsymbol{\Sigma} = \{\boldsymbol{\Sigma}_1, ..., \boldsymbol{\Sigma}_n\}$ is the variance of the noise. It is noted that alternative (\emph{e.g.}, multiplicative) noise models can be used \cite{Kaipio2006StatisticalProblems}. Furthermore, we note that, to avoid arithmetic underflow  problems, in our implementation we consider the log-likelihood function
\begin{equation}
    l(\boldsymbol{\theta}|\boldsymbol{y}) = \ln\left(L(\boldsymbol{\theta}|\boldsymbol{y})\right),
\end{equation}
instead of the likelihood function itself.

The prior density in equation~\eqref{eq:bayesrule} is based on the acquired knowledge prior to obtaining the experimental data. This term is based on experts' knowledge, intuition, or previous experiments. In uncertainty propagation, we use the prior as the parametric uncertainty which is propagated through the model. If no relevant information is known about a parameter, one typically uses an uninformative prior, for which often a uniform density is used \cite{Lambert2017AStatistics}. If sufficient and convincing information is obtained for a parameter's value, the prior is informative with a relatively small uncertainty. If the uncertainty in the prior is relatively small in comparison to the experimental noise, the posterior will strongly be influenced by the prior, and \emph{vice versa}.

On account of the in general high dimensionality of the parameter domain, the evidence can be computationally demanding to evaluate since it involves an integral over all parameter values. There are several ways to evaluate it, where the method depends on the number of parameters in the systems. In special cases, the term can be evaluated analytically. For low-dimensional problems, \emph{i.e., $p \lessapprox 4$}, one typically uses quadrature techniques \cite{Smith2013UncertaintyApplications}. In a moderate number of dimensions, (adaptive) sparse grids are used \cite{Gerstner1998NumericalGrids}. For high-dimensional problems, Monte Carlo methods, such as Markov chain Monte Carlo (MCMC) algorithms \cite{Brooks1998MarkovApplication}, are commonly used. In these samplers, only the probability ratio of two subsequent steps in the chain is needed, making the evaluation of the evidence unnecessary.

\section{Squeeze flow setup and experiment}
\label{sec:setup}

In this section we specify all relevant aspects of the tailored experimental setup and experiment. 

\subsection{Fluids}
The squeeze flow experiments are performed using two types of fluids: glycerol and a Polyvinylpyrrolidone (PVP) solution. The former is in liquid form obtained from VWR Chemicals with purity $\geq$99.5\%. It has been exposed to air for two days to be saturated with water absorbed from the air. The PVP is obtained in powdered form (average molecular weight of 360 kg/mol) from Sigma Aldrich, which is dissolved in purified water. A 23 wt.\% PVP aqueous solution is obtained by stirring the mixture for three days. 

\subsection{Instrumentation}
\begin{figure}
     \centering
     \subfloat[]{\includegraphics[width=0.23\textwidth]{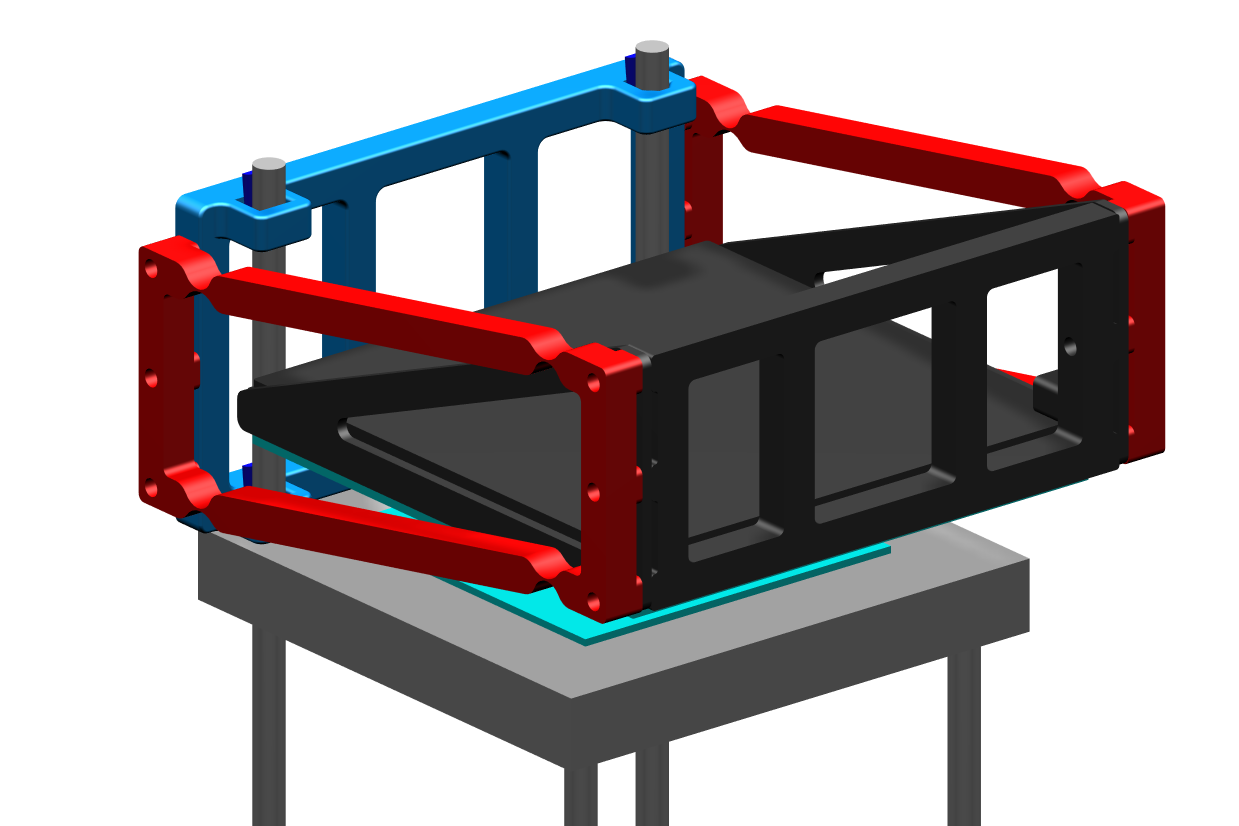}\label{fig:setup_initial}}
     \hfill
     \subfloat[]{\includegraphics[width=0.23\textwidth]{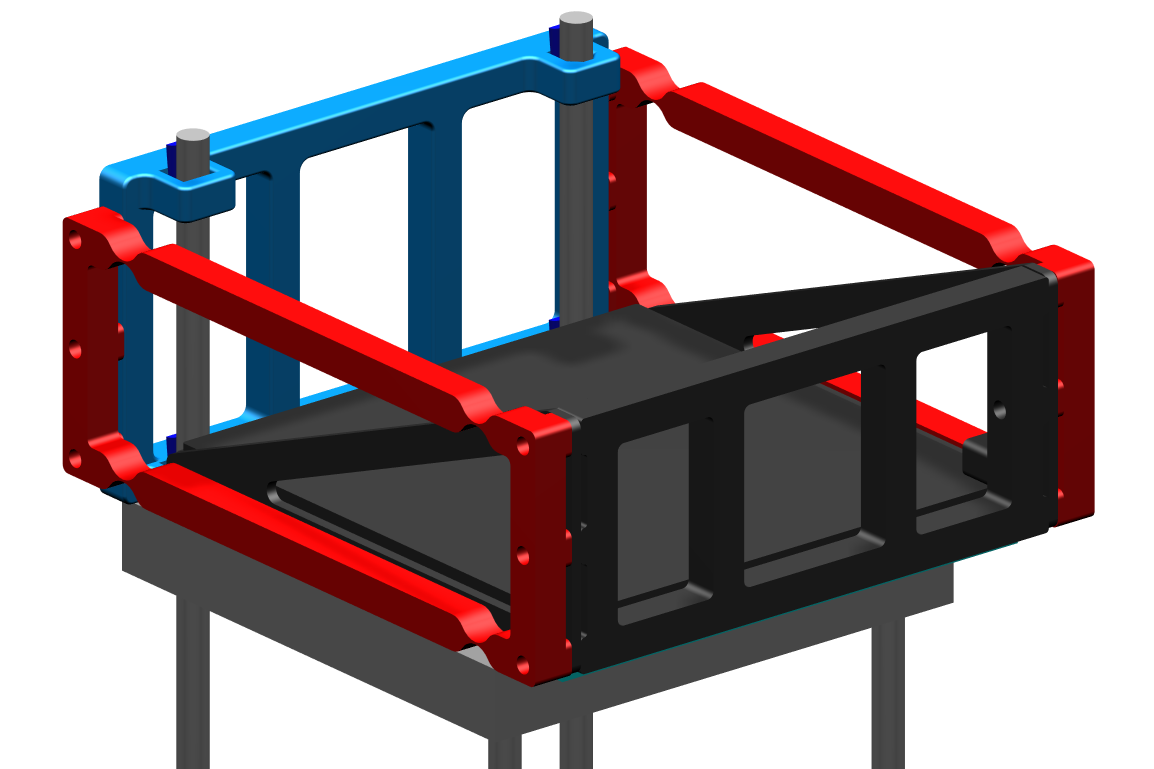}\label{fig:setup_final}}     
    \caption{Tailored experimental setup, with an aluminium frame (grey), moving body (black), leafsprings (red) and PMMA plates (cyan): (a) starting position, (b) fluid compression position.}\label{fig:setup_schem}
\end{figure}

To have control over all aspects of the experiment, we developed a tailored setup (\autoref{fig:setup_schem}) to perform the squeeze flow measurements. The setup consists of an aluminium frame and a 3D-printed parallel-beam construction. Furthermore, two Polymethylmethacrylaat (PMMA) plates are used, in between which the fluid is compressed. One plate is attached to the 3D-printed construction and the other one to the aluminium frame. The positions of the setup before and during an experiment are shown in \autoref{fig:setup_initial} and \autoref{fig:setup_final}, respectively. The red beams act as leaf springs, ensuring the parallel motion of the PMMA plates. To capture the fluid flow behavior, a Nikon Coolpix W300 camera is used\footnote{The camera settings are given in the supplementary information (Camerasettings.pdf).}. The 3D printed parts are made from Polylactic Acid (PLA) and have been developed with the Ultimaker S5, which is a 3D printer using Fused deposition modeling (FDM). Each part is designed in the software SIEMENS NX and transmitted to the software Ultimaker Cura to enable printing\footnote{The printer settings in Ultimaker Cura are provided in the supplementary information (Printersettings.pdf). }. 

\subsection{Procedure}
In our experiments, we extract glycerol using a syringe. The PVP solution is extracted using a small spoon because the fluid has a too high viscosity to be extracted by a syringe. The bottom PMMA plate is removed from the setup to deposit the material on it. When reassembling the setup, a release mechanism is placed between the top and bottom plate to ensure a controlled height of the fluid layer. We ensure that the fluid touches both plates before the start of the experiment to properly determine the initial height. The camera is remotely focused on the sample using the SnapBridge app. When the sample is in focus, the camera can start recording and the release mechanism is triggered. The motion of the fluid is captured for a duration of five minutes. After measuring, we use water to clean the plates and dry them before positioning a new sample. Each experimental case (see \autoref{tab:expcases}) is performed ten times.

We tested several cases for both materials, where we varied the applied force and fluid volume. The cases are listed in \autoref{tab:ExpCaseGly} and \autoref{tab:ExpCasePVP} for glycerol and PVP solution, respectively. The applied force is adjusted by placing additional weights on the top plate. The amount of additional weight is denoted by $F_{\mathrm{add}}$. Due to the high viscosity fluids being used, determining the deposited volume is challenging. The intended volume for glycerol is indicated by markings on the pipette, represented by $V_\mathrm{a}$ and $V_\mathrm{b}$, where $V_\mathrm{a}$ $<$ $V_\mathrm{b}$. For the PVP solution, the intended volume is indicated by drawing two differently sized circles on the PMMA plate denoted by $V_\mathrm{c}$ and $V_\mathrm{d}$, where $V_\mathrm{c}$ $<$ $V_\mathrm{d}$. The values of these fluid volumes and their uncertainty are discussed in Section \ref{sec:uncertainties}.
\begin{table}
   \centering
           \captionsetup[subtable]{position = below}
          \captionsetup[table]{position=top}
           \caption{Squeeze flow experimental cases: (a) using glycerol, where $V_\mathrm{a}$ $<$ $V_\mathrm{b}$ (b) using PVP solution, where $V_\mathrm{c}$ $<$ $V_\mathrm{d}$. See Section~\ref{sec:uncertainties} for details.}
           \label{tab:expcases}
           \begin{subtable}{0.5\linewidth}
               \centering
               \begin{tabularx}{0.95\columnwidth}{XXX}
                   \toprule
                   Case & $F_{\mathrm{add}}$ [kg] & $V$ [-] \\ \midrule
                   \RNum{1} & 0 & $V_\mathrm{a}$ \\ 
                   \RNum{2} & 0.25 & $V_\mathrm{a}$ \\ 
                   \RNum{3} & 0.5 & $V_\mathrm{a}$ \\ 
                   \RNum{4} & 0.25 & $V_\mathrm{b}$ \\ \bottomrule
               \end{tabularx}
               \caption{}
               \label{tab:ExpCaseGly}
           \end{subtable}%
           \begin{subtable}{0.5\linewidth}
               \centering
               \begin{tabularx}{0.95\columnwidth}{XXX}
                   \toprule
                   Case & $F_{\mathrm{add}}$ [kg] & $V$ [-] \\ \midrule
                   \RNum{5} & 0 & $V_\mathrm{c}$ \\
                   \RNum{6} & 0.5 & $V_\mathrm{c}$ \\ 
                   \RNum{7} & 0 & $V_\mathrm{d}$ \\ 
                   \RNum{8} & 0.5 & $V_\mathrm{d}$ \\ \bottomrule
               \end{tabularx}
               \caption{}
                 \label{tab:ExpCasePVP}
           \end{subtable}
\end{table}
\subsection{Processing}
\begin{figure}
     \centering
     \subfloat{\fbox{\includegraphics[width=0.22\textwidth]{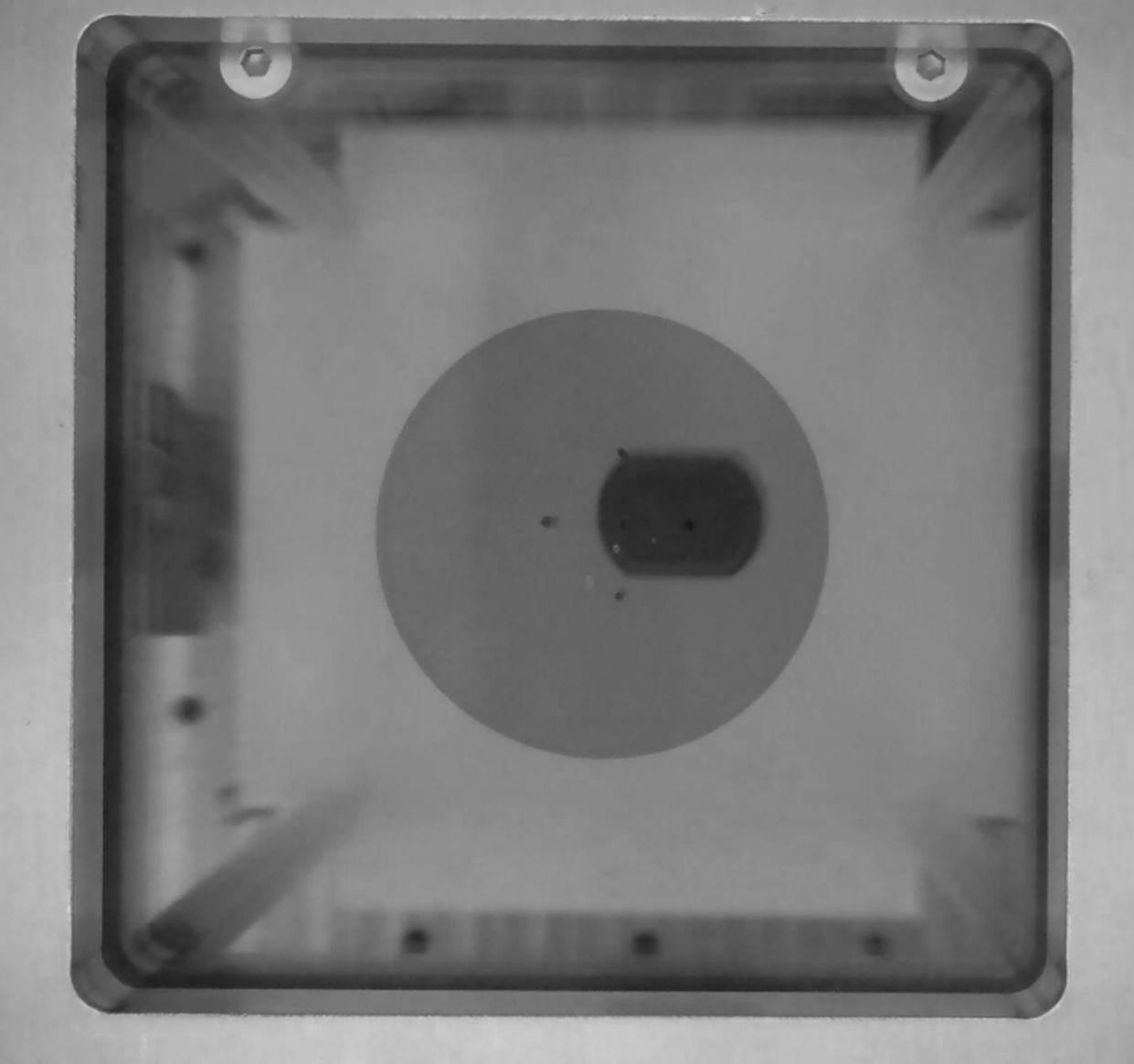}}\label{chSetup_Im}}      
     \hfill
     \subfloat{\fbox{\includegraphics[width=0.22\textwidth]{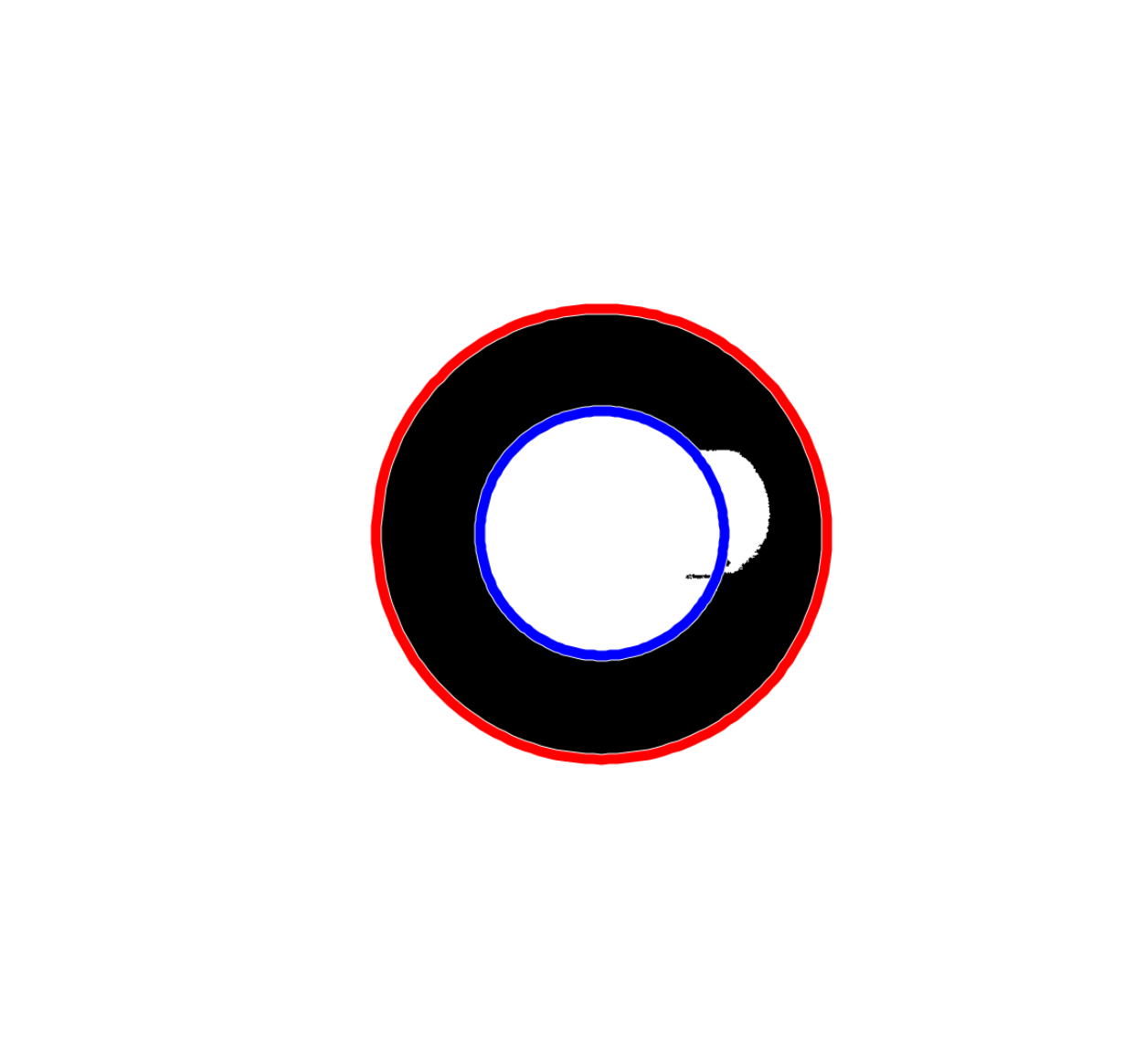}}\label{chSetup_BwIm}}
    \caption{Image processing for case \RNum{6} at t = 9.14\,s. (left) Original image retrieved from the camera. (right) Binary image created by subtracting the current frame from the initial frame and using Otsu's method \cite{Bangare2015ReviewingThresholding} to determine the threshold value. Using the circle Hough Transform \cite{Yuen1990ComparativeFinding}, the initial fluid front (blue) and the current fluid front (red) are determined, assuming a circular shape. Subsequently, the radius of the fluid layer is evaluated.}\label{chSetup_imageproc}
\end{figure}

We analyze the experimental data by extracting the radius of the fluid layer using image processing, as illustrated in \autoref{chSetup_imageproc}. The frames are retrieved with a frame rate of 60\,Hz. Due to the delay between the recording of the camera and the start of the measurement, the initial frame is determined by comparing the frames up until the fluid is set in motion.

\section{Squeeze flow model}
\label{sec:model}

\begin{figure}
     \centering
     \subfloat[]{\includegraphics[width=0.23\textwidth]{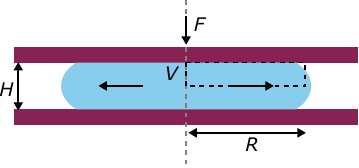}\label{fig:SqueezeflowEXPLAIN}}
     \hfill
     \subfloat[]{\includegraphics[width=0.23\textwidth]{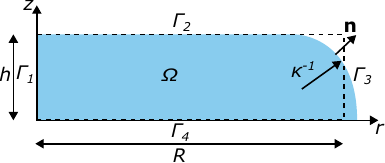}\label{fig:Squeeze_domain}}
    \caption{Overview of the squeeze flow model, where two parallel plates (dark-red) compress a fluid with volume $V$ (light blue) by applying a force $F$. The distance between the parallel plates $H=2h$ and the radius of the fluid layer $R$ evolve over time. The model domain is denoted by the dashed black box and cylindrical coordinates are used for solving the problem.}
     \label{fig:SqueezeFlow}
\end{figure}

In this section we present the considered squeeze flow model. A schematic of the model is provided in \autoref{fig:SqueezeFlow}. The primary quantity of interest for our model is the radius of the fluid layer over time, $R(t)$. To model the evolution of the radius, at any given moment in time, we assume the problem to be isothermal. Furthermore, we assume negligible inertial forces and the fluid to be incompressible. As a starting point, we consider the mass and momentum balance given by
\begin{subequations}\label{balancelaws}
    \begin{align}
        \nabla \cdot \boldsymbol{v} & = 0, \label{ch3_mass_bal_1}\\
        \nabla p - \nabla \cdot \boldsymbol{\tau} & = \mathbf{0}, \label{ch3_mom_bal_1}
    \end{align}
\end{subequations}
where $\boldsymbol{v}$ and $p$ are the velocity and pressure field, respectively, both defined over the volumetric fluid domain $\Omega(t)$. The extra stress $\boldsymbol{\tau}$ is related to these fields through a constitutive rheological model, which will be discussed in detail below. Note that the domain $\Omega(t)$ is time-dependent, being directly related to the radius $R(t)$. In our model, the motion of the fluid front is governed directly by the velocity field at the fluid-air interface, $\Gamma_3(t)$.

We assume the solution to be axisymmetric, meaning that we assume independence of the circumferential coordinate. Assuming the fluid layer to be thin in comparison to the radius, \emph{i.e.}, $H \ll R$, we use lubrication theory \citep{Szeri2010FluidLubrication} to dimensionally reduce the mass and momentum balance \eqref{balancelaws} (derivation presented in \ref{app:scalemommass}) to
\begin{subequations}
\begin{align}
    \frac{1}{r}\frac{\partial}{\partial r}\left(rv_r\right) - \frac{\partial v_z}{\partial z} & = 0, \label{eq:massBal} \\
    \frac{\partial}{\partial z}\tau_{rz} & = \frac{\partial p}{\partial r}\label{eq:mombal},
\end{align}
\end{subequations}
where $v_r$ is the $r$-component of the velocity vector $\boldsymbol{v}$, $v_z$ the $z$-component of $\boldsymbol{v}$ and $\tau_{rz}$ is the $rz$-component of the stress tensor $\boldsymbol{\tau}$. All other components of the velocity vector and extra stress tensor are neglected.

To solve these conservation equations on the evolving domain, we develop a model using the simulation domain given in \autoref{fig:Squeeze_domain}, where the boundaries are denoted by $\Gamma_2$, $\Gamma_3$ and $\Gamma_4$ and the axisymmetric axis denoted by $\Gamma_1$. At $\Gamma_2$ we have a no-slip condition between the fluid and the plates, \emph{i.e.}, $v_r(r,h)=0$. Furthermore, the vertical velocity $v_z$ equals half the velocity of the upper plate $\dot{H}$, \emph{i.e.}, $v_z(r,h)=\frac{1}{2}\dot{H}=\dot{h}$. An additional condition is set at $\Gamma_2$ stating that the pressure integrated over the top plate equals the applied force, \emph{i.e.}, $\int_0^R p(r,t) 2\pi r\, {\rm{d}} r=F$. Note that the atmospheric pressure is omitted, which is justifiable due to the fluid incompressibilty assumption. At the axisymmetry axis, $\Gamma_1$, the radial velocity equals zero, \emph{i.e.}, $v_r(0,z)=0$. At the horizontal symmetry plane, $\Gamma_4$, the shear stress is equal to zero, \emph{i.e.}, $\tau_{rz}(z,r,t)=0$. At the free boundary $\Gamma_3$ we have a traction boundary condition where the normal traction equals the Laplace pressure, \emph{i.e.}, $p(R(t),z)=\Delta p$, with
\begin{equation}\label{eq:Laplace}
   \Delta p= \gamma \nabla_{\rm{s}} \cdot \mathbf{n} = 2 \gamma \kappa,
\end{equation}
where $\gamma$ is the surface tension, $\nabla_{\rm{s}}$ is the surface gradient operator defined by $\nabla_{\rm{s}}=(\mathbf{I}-\mathbf{n} \otimes \mathbf{n})\cdot \nabla$, where $\mathbf{I}$ is the unit tensor and $\mathbf{n}$ is the normal to the interface, and $\kappa$ the curvature. The curvature around the $z$-axis is assumed to be significantly smaller than the curvature between the plates ($\kappa^{-1}$). The former curvature is therefore neglected. 

\begin{figure}
    \centering
    \includegraphics[width=0.5\columnwidth]{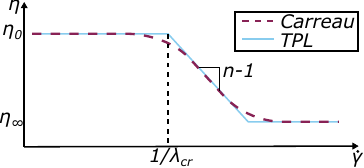}
    \caption{The Carreau model and the truncated power law model in simple shear (with shear rate $\dot{\gamma}$) describe the Newtonian regimes (\emph{i.e.}, at the zero shear rate viscosity and at the high shear rate viscosity) and the power law regime similarly. The transition between these regimes is smooth for a Carreau model and sharp for a truncated power law model. Note that the axes are assumed logarithmic in this schematic.}
    \label{fig:SchemCarreauTPL}
\end{figure}

We herein consider two constitutive models to be used in combination with the squeeze flow model: a Newtonian model and a generalized Newtonian model. The former describes fluids where the viscosity is independent of the shear rate, while the latter is able to mimic fluids with combined Newtonian and shear thinning behavior.
A commonly used model for shear thinning is the Carreau model\,\cite{Bird1987DynamicsMechanics}, which combines Newtonian behavior at high and low shear rates, and introduces a shear thinning behavior for intermediate shear rates. However, we
 use a truncated power law (TPL) model instead of a Carreau model because the squeeze flow model can then be solved semi-analytically, while still describing similar flow behavior, thereby saving computational time. The truncated power law model is illustrated in \autoref{fig:SchemCarreauTPL} to show the main difference with the Carreau model.

In the remainder of this section the squeeze flow model will be elaborated for the two considered constitutive models. It is noted that formally the Newtonian model is a special case of the truncated power law model. For the clarity of the derivation, we however opt to first consider the Newtonian model as a separate case.

\subsection{Newtonian model}
To set the scene for the derivation of the truncated power law model in Section \ref{subsec:TPL}, we briefly review the fundamental steps in the derivation for the Newtonian case, including the Laplace pressure boundary condition. We note that in the absence of the Laplace pressure, the result is well established in the literature, see, \emph{e.g.}, \cite{Engmann2005SqueezeReview}.

The first step in the derivation of the expression for the radius (and height) over time is to use the momentum balance \eqref{eq:mombal} to acquire the through-the-thickness velocity profile as a function of the pressure gradient $\partial p/ \partial r$ and the rheological parameters. For the Newtonian case, the viscosity is a constant denoted by $\eta_\mathrm{N}$, thus $\tau_{rz}$ simplifies to
\begin{equation}
    \tau_{rz} = 2\eta_\mathrm{N} D_{rz} = \eta_\mathrm{N} \frac{\partial v_r}{\partial z},
\end{equation}
where $D_{rz}$ is the $rz$-component of the rate-of-deformation tensor $\textbf{\emph{D}}$. By substitution of this expression in the momentum balance \eqref{eq:mombal}, we obtain the radial velocity 
\begin{equation}
     v_r(r,z) = \frac{1}{2\eta_\mathrm{N}} \frac{\partial p}{\partial r}\left(z^2 - h^2\right),
\end{equation}
where use has been made of the boundary conditions $v_r(r,h)=0$ and $\tau_{rz}(r,0)=0$.

To obtain the pressure field over time, the obtained velocity profile is substituted in the thickness-integrated mass balance \eqref{eq:massBal} as
\begin{equation}
   \int_{0}^{h}\frac{1}{r}\frac{\partial}{\partial r}\left(r v_r\right) \, {\rm{d}} z + \dot{h} = 0, 
\end{equation}
where use is made of the boundary conditions $v_z(r,0)=0$ and $v_z(r,h)=\dot{h}$. Evaluation of the integrated mass balance then yields
\begin{equation}\label{eq:pressurefieldN}
    p(r) = \frac{3 \eta_\mathrm{N} \dot{h}}{ 4h^3}\left(r^2-R^2\right) + \Delta p,
\end{equation}
where we have used $\frac{\partial p }{\partial r}(r$\,=\,$0)$\,=\,$0$ and $p(r=R)$\,=\,$\Delta p$. 

Note that the pressure field \eqref{eq:pressurefieldN} is a function of the speed of the top plate. Since our model is force-driven, the pressure field must be expressed in terms of the applied force instead of the plate's speed. To this end, we use the additional condition $\int_0^Rp(r)2\pi r {\rm{d}}r$\,=\,$F$ to obtain
\begin{equation}\label{eq:hdotN}
   \dot{h} = \frac{4}{3} \left( -\frac{2 Fh^3}{\pi \eta_\mathrm{N} R^4} + \frac{4 \kappa \gamma h^3}{\eta_\mathrm{N} R^2} \right),  
\end{equation}
from where we obtain the height and radius of the fluid layer over time using the forward Euler method \cite{Biswas2013AReview} and the initial condition $h(t$\,=\,$0$)\,=\,$h_0$. 
\begin{remark}[Zero Laplace pressure special case]
The squeeze flow model can be solved analytically in the special case where we assume the Laplace pressure to be equal to zero \cite{Engmann2005SqueezeReview}. In that case one obtains from \eqref{eq:hdotN} that the height and radius evolve as
\begin{subequations}
    \begin{align}
        H & = H_0\left(1+\frac{8 H_0^2 F t}{3 \pi \eta_\mathrm{N} R_0^4}\right)^{-1/4}, \\
        R & = \sqrt{\frac{V}{\pi H_0} \left(1+\frac{8 H_0^2 F t}{3 \pi \eta_\mathrm{N} R_0^4}\right)^{1/4}} = R_0 \left( 1 + \frac{8 F V^2}{3 \eta_N \pi^3 R_0^8} t \right)^{1/8}. 
    \end{align}
\end{subequations}
\end{remark}

\subsection{Truncated power law model}\label{subsec:TPL}
\begin{figure}
     \centering
     \subfloat[]{\includegraphics[width=0.7\columnwidth]{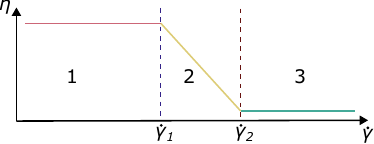}\label{fig:TPLregiongraph}}
     \hfill
     \subfloat[]{\includegraphics[width=0.7\columnwidth]{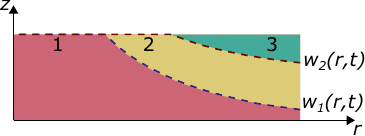}\label{fig:Tplregionsqueeze}}
    \caption{Schematic of the characterization of the three possible flow regions, where we have three different constitutive models: 1) (pink) $\eta=\eta_0$, 2) (yellow) $\eta =\eta_0 \left( \lambda_{\mathrm{cr}}|\dot{\gamma}|\right)^{n-1}$ and 3) (green) $\eta=\eta_\infty$. (a) The division of the regions in simple shear (note that the axes are assumed logarithmic in this
    schematic), (b) The division of the regions in a squeeze flow experiment at a fixed moment in time. The heights of the interfaces between the regions are denoted by $w_1$ and $w_2$, which are both a function of $r$. These interfaces are shown in dashed purple and dark red, respectively.} 
     \label{fig:TPLregions}
\end{figure}

The truncated power law model describes Newtonian and shear thinning flow behavior, as illustrated in \autoref{fig:TPLregiongraph}. Note that during our experiments we do not reach the third region. However, to improve numerical stability we also consider this region. In other research the truncated power law model has been used in \emph{e.g.} \cite{Lavrov2015FlowApplications, Lee2023ModellingSurfaces}. However, Lee \emph{et al.} \cite{Lee2023ModellingSurfaces} have modeled two regions instead of three and Lavrov \cite{Lavrov2015FlowApplications} has modeled three regions with the pressure gradient as pre-knowledge. Therefore, to the best of our knowledge, the three-region case with arbitrary pressure gradient, which we consider here, has not been reported.

Due to the lubrication assumption, the squeeze flow is rheometric, \emph{i.e.},~it can be considered a point-wise simple shear flow with absolute shear rate $|\dot{\gamma}|=\left| \frac{\partial v_r}{\partial z} \right|$ (\ref{app:scaleshear}). The corresponding mathematical model for the viscosity then reads
\begin{equation}
    \eta \left(|\dot{\gamma}|\right) = \begin{cases} \eta_0 & 0\leq |\dot{\gamma}| \leq \dot{\gamma}_1 \\
    \eta_0 \left( \lambda_{\mathrm{cr}} \left| \dot{\gamma} \right| \right)^{n-1} & \dot{\gamma}_1 < |\dot{\gamma}| < \dot{\gamma}_2 \\
    \eta_\infty & \dot{\gamma}_2 \leq |\dot{\gamma}| < \infty
    \end{cases}
\end{equation}
where $\eta_0$, $\lambda_{\mathrm{cr}}$, $n$ and $\eta_\infty$ are the viscosity at the zero shear rate plateau, the inverse of the shear rate corresponding to the transition between region 1 and 2, the power index and the viscosity at infinite shear rate. Note that for notational convenience we have rewritten the power law constitutive model from its more common representation $\eta = K \left| \dot{\gamma} \right|^{n-1}$, where $K$ is the flow consistency index, by introducing $K=\eta_0\lambda_{\rm{cr}}^{n-1}$.

During a squeeze flow experiment, the maximum shear rate decreases from the high shear rate plateau to the low shear rate plateau over time. Furthermore, at a fixed moment in time, the shear rate near the top plate is higher than the shear rate in the middle of the fluid layer, as visualized in \autoref{fig:Tplregionsqueeze}. Therefore, it is essential that the three possible regions of the flow behavior are incorporated in the squeeze flow model.

As for the Newtonian case, the first step in our derivation of the radius over time is to solve the momentum balance to attain the velocity profile as a function of the pressure gradient and rheological parameters. The fundamental complication in the case of the truncated power law is that the profile is subdivided in three regions, parameterized by the heights $w_1$ and $w_2$, where the interface positions themselves depend on the pressure drop and rheological parameters. In our derivation, we commence with solving the momentum balance in each of the three regions, after which we use the continuity of traction and velocity across the interfaces to obtain the velocity profile with assumed values for the interface heights. We then determine the analytical profile by using the continuity of the viscosity to derive expressions for the interface positions.

\begin{figure}
    \centering    \includegraphics[width=\columnwidth]{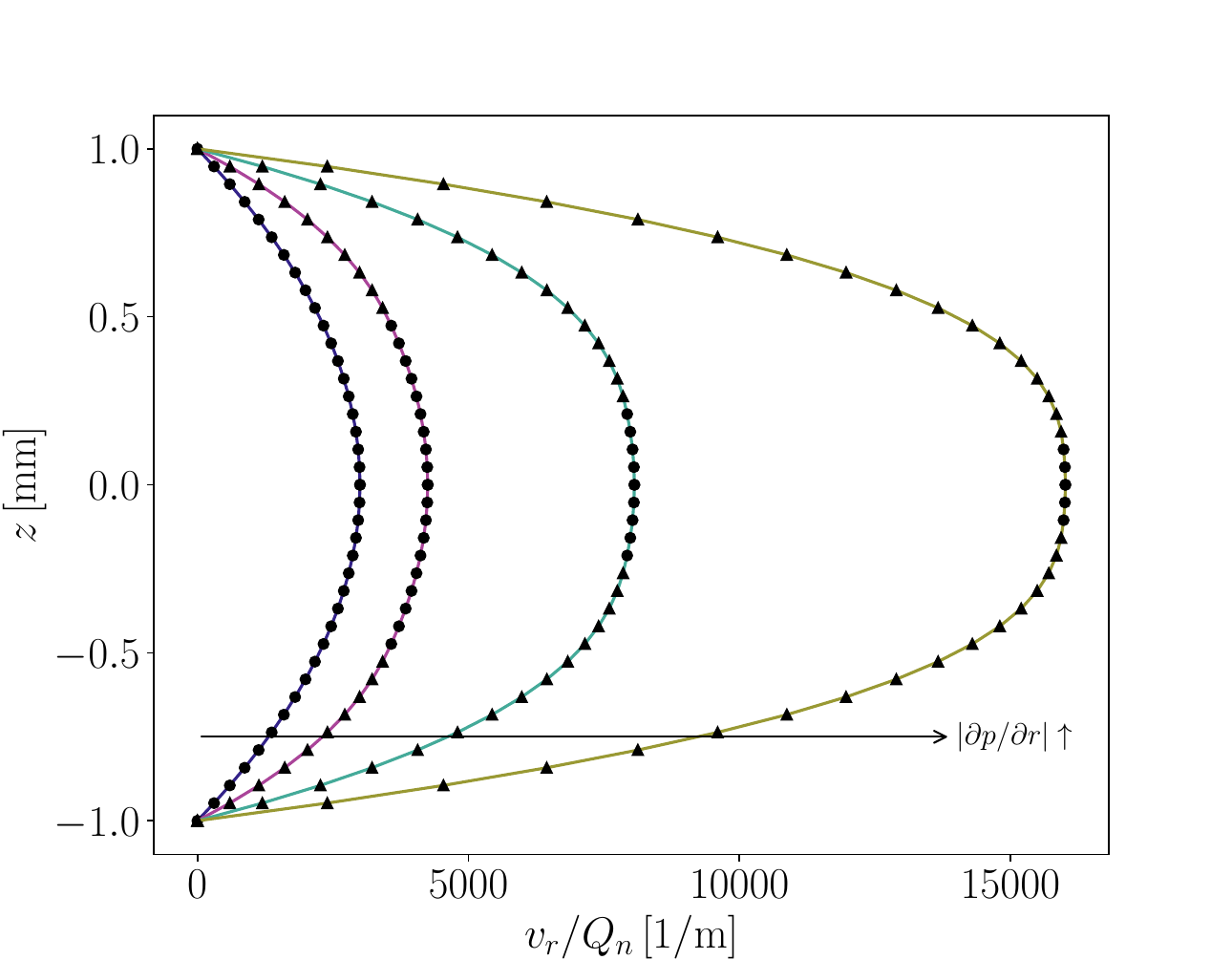}
    \caption{Velocity profiles of $\frac{\partial p}{\partial r}=\left[\frac{\partial p}{\partial r}\big|_{\lambda_{\mathrm{cr}}}, 2\frac{\partial p}{\partial r}\big|_{\lambda_{\mathrm{cr}}}, 4\frac{\partial p}{\partial r}\big|_{\lambda_{\mathrm{cr}}}, 8\frac{\partial p}{\partial r}\big|_{\lambda_{\mathrm{cr}}}\right]$, where $\frac{\partial p}{\partial r}\big|_{\lambda_{\mathrm{cr}}}$ is the pressure gradient that yields a maximum shear rate of $1/\lambda_\mathrm{cr}$. The velocity is normalized by the Newtonian flux for the corresponding $\frac{\partial p}{\partial r}$ values. The black triangles and circles denote the shear thinning and Newtonian behavior, respectively. As the pressure gradient increases, the shear thinning behavior increase throughout the velocity profile. }
    \label{fig:velprofile}
\end{figure}

To start, the momentum balance is solved by implementing the relevant constitutive models per region. The extra stress is then defined as 
\begin{equation}\label{eq:tauconstmodel}
    \tau_{rz} = \begin{cases} \eta_0 \frac{\partial v_r}{\partial z} & 0\leq z \leq w_1 \\
   \eta_0 \left( \lambda_{\mathrm{cr}} \left| \frac{\partial v_r}{\partial z} \right| \right)^{n-1} \frac{\partial v_r}{\partial z}, & w_1 < z < w_2 \\
   \eta_\infty \frac{\partial v_r}{\partial z} & w_2 \leq z \leq h \end{cases}.
\end{equation}
Using the boundary condition $\tau_{rz}(z$\,=\,$0)$\,=\,0 and traction continuity at the interfaces $z$\,=\,$w_1$ and $z$\,=\,$w_2$, we obtain the stress from the momentum balance as
\begin{equation}\label{eq:taudpdr}
    \tau_{rz} = \frac{\partial p}{\partial r} z.
\end{equation}
To obtain the velocity profile through-the-thickness,
\begin{equation}
    v_r = \begin{cases}
        v_{r,1} & 0 \leq z \leq w_1 \\
        v_{r,2} & w_1 < z < w_2 \\
        v_{r,3} & w_2 \leq z \leq h
    \end{cases},
\end{equation}
we match the expressions provided in equation \eqref{eq:tauconstmodel} and \eqref{eq:taudpdr}. Making use of the boundary condition $v_r(z$\,=\,$h)$\,=\,0 and the continuity condition of $v_r$ at the interfaces gives
\begin{subequations}
    \begin{align}
    \begin{split}
        v_{r,1} & = \frac{1}{2}\frac{1}{\eta_0}\frac{\partial p}{\partial r} z^2 \\
        & \quad + \frac{\partial p}{\partial r} \left( \frac{t_{cr}^{1-n}}{\eta_0} \right)^\frac{1}{n} \left| \frac{\partial p}{\partial r} \right|^\frac{1-n}{n} \frac{n}{n+1} \left( w_1^\frac{n+1}{n} - w_2^\frac{n+1}{n} \right) \\
        & \quad + \frac{1}{2} \frac{1}{\eta_0} \frac{\partial p}{\partial r} \left( \frac{\eta_0}{\eta_\infty} w_2^2 - \frac{\eta_0}{\eta_\infty} h^2 - w_1^2 \right),
    \end{split} \\
    \begin{split}
        v_{r,2} & = \frac{\partial p}{\partial r} \left( \frac{t_{cr}^{1-n}}{\eta_0} \right)^\frac{1}{n} \left| \frac{\partial p}{\partial r} \right|^\frac{1-n}{n} \frac{n}{n+1} \left( z^\frac{n+1}{n} - w_2^\frac{n+1}{n} \right) \\
        & \quad + \frac{1}{2} \frac{1}{\eta_\infty} \frac{\partial p}{\partial r} \left( w_2^2 - h^2 \right),
    \end{split} \\
    v_{r,3} & = \frac{1}{2} \frac{1}{\eta_\infty} \frac{\partial p}{\partial r} \left( z^2 - h^2 \right). 
    \end{align}%
\end{subequations}
The velocity profile is visualized in \autoref{fig:velprofile} for numerous pressure gradients, yielding flows with shear rates in the Newtonian plateau and power law regimes. The velocity profile flattens as the gradient of the pressure increases, which is due to the shear thinning behavior of the fluid. 

Next, we determine the total flux by the addition of the fluxes per region as
\begin{equation}
    Q = Q_1 + Q_2 + Q_3,
    \label{eq:TPLtotalflux}
\end{equation}
where the regional fluxes are determined by integrating the expressions for $v_r$ over the specified height per region, given by 
\begin{subequations}
    \begin{align}
        \begin{split}
            Q_1 & = \frac{1}{\eta_\infty} \frac{\partial p}{\partial r} \left( w_2^2 w_1 - h^2 w_1 - \frac{2}{3}\frac{\eta_\infty}{\eta_0} w_1^3 \right) \\
                & \quad + 2 \frac{\partial p}{\partial r} \left( \frac{\lambda_{\mathrm{cr}}^{1-n}}{\eta_0} \right)^\frac{1}{n} \left| \frac{\partial p}{\partial r} \right|^\frac{1-n}{n} \frac{n}{n+1} \left( w_1^\frac{n+1}{n} - w_2^\frac{n+1}{n} \right) w_1,
        \end{split} \\
        \begin{split}
            Q_2 & = 2 \frac{\partial p}{\partial r} \left( \frac{\lambda_{\mathrm{cr}}^{1-n}}{\eta_0} \right)^\frac{1}{n} \left| \frac{\partial p}{\partial r} \right|^\frac{1-n}{n} \frac{n}{n+1} \\
                & \quad \cdot \left( - \frac{n+1}{2n+1} w_2^\frac{2n+1}{n} - \frac{n}{2n+1} w_1^\frac{2n+1}{n} - w_2^\frac{n+1}{n} w_1 \right) \\ 
                & \quad + \frac{1}{\eta_0} \frac{\partial p}{\partial r} \left( w_2^2 - h^2 \right) \left( w_2 - w_1 \right),   
        \end{split} \\
            Q_3 & = \frac{1}{\eta_\infty} \frac{\partial p}{\partial r} \left( -\frac{2}{3} h^3 - \frac{1}{3} w_2^3 + h^2 w_2 \right).
    \end{align}    
    \label{eq:TPLfluxes}%
\end{subequations}
We incorporate the flux $Q$ in the conservation of mass as
\begin{equation}\label{eq:TPLstrong}
    -\frac{\partial}{\partial r} \left(Q r\right) = 2 r \dot{h},
\end{equation}
which is a non-linear differential equation in time, where the non-linearity stems from the dependence of the interface positions, $w_1(r)$ and $w_2(r)$, on the solution. Using the viscosity continuity condition, this dependence can be expressed as 
\begin{subequations}
    \begin{align}
    w_1 & = \frac{\eta_0}{\lambda_{\mathrm{cr}}} \left| \frac{\partial p}{\partial r} \right|^{-1}, \\
    w_2 & = \eta_0 \lambda_{\mathrm{cr}}^{-1} \left( \frac{\eta_\infty}{\eta_0} \right)^\frac{n}{n-1} \left| \frac{\partial p}{\partial r} \right|^{-1}.
\end{align}
\label{eq:TPLws}%
\end{subequations}
To integrate \eqref{eq:TPLstrong} in time, we have developed a semi-analytical non-linear time-integrator based on fixed point iterations. The typical runtime of our model is $t_{\mathrm{sim}}=2$\,s for a simulation time of $t=350$\,s. We refer the reader to \ref{sec:APPsolver} for details about the solver.

\begin{remark}[Limiting cases of the truncated power law]
    When $| \frac{\partial p}{\partial r} | \leq \frac{\eta_0}{h \lambda_\mathrm{cr}}$ it follows that $w_1=w_2=h$ and we get $\eta=\eta_0$ and the solutions~\eqref{eq:pressurefieldN} and \eqref{eq:hdotN}. When $| \frac{\partial p}{\partial r} | \rightarrow \infty$ it follows that $w_1=w_2 \rightarrow 0$ and we get $\eta=\eta_\infty$ and the solutions~\eqref{eq:pressurefieldN} and \eqref{eq:hdotN}, but with $\eta_\infty$ instead of $\eta_0$. Note that in these cases, the solution can be expressed analytically, but that this is not possible in the general case.
\end{remark}

\begin{remark}\label{rmk:logtransform}
    Due to physical considerations, the rheological model parameters  $\eta_\mathrm{N}$, $\eta_0$, $\eta_\infty$ and $\lambda_{\mathrm{cr}}$ and squeeze flow parameters $V$ and $R_0$ must be non-negative. To avoid these parameters from becoming negative in the sampling method introduced in Section \ref{sec:uncertainties}, a log-transformation is applied. This is done by introducing the transformation $\boldsymbol{\theta}$ = $\exp$($\hat{\boldsymbol{\theta}}$), where the model parameters are denoted by $\boldsymbol{\theta}$, and assigning a distribution to the transformed parameters $\hat{\boldsymbol{\theta}}$.
\end{remark}

\section{Characterization of parametric uncertainties}
\label{sec:uncertainties}

The input parameters used to determine the quantity of interest can either be probabilistic or deterministic. Parameters can be probabilistic due to a lack-of-knowledge of a deterministic value or due to physical randomness in the system.

As defined in Section~\ref{sec:model}, the input parameters of the squeeze flow model are the fluid volume $V$, the applied force $F$, the initial radius of the fluid $R_0$ and the rheological parameters, \emph{i.e.}, $\eta$ for the Newtonian model and $\eta_0$, $\eta_\infty$, $n$ and $\lambda_\text{cr}$ for the truncated power law. 
The squeeze flow experiment introduces additional uncertainties. These include the refraction of the camera lens and PMMA plates, the pixel-to-length ratio of the squeeze flow images and the uncertainties associated with the circular fit of the fluid front. We assume that these additional parameters are deterministic, because our experience suggests that the uncertainty of the model prediction is dominated by the stochastic parameters mentioned above (\emph{i.e.}, the squeeze flow model parameters).

The parametric uncertainties for $R_0$, $V$ and $F$ are determined by performing dedicated experiments for each of the parameters. Each experiment is repeated ten times, from which a parametric distribution is defined. Because it is difficult to perform independent measurements for each of the truncated power law parameters, these constitutive model parameters are determined by Bayesian inference.

\subsection{Sampling method}
In our Bayesian inference problems, we use a Markov chain Monte Carlo (MCMC) sampling method, which draws a number of realizations from the posterior distribution by evaluating the prior and likelihood for any particular parameter value. As the sample size (\emph{i.e.}, the number of realizations) is increased, the sampled distributions can be expected to move closer to the posterior distribution. In the considered Markov chain Monte Carlo method, the next step in the chain of realizations is only dependent on the latest parameter value. Multiple chains (or walkers) are used to robustly explore the parameter space, circumventing the possibility of only sampling a local region of high probability. In this work, the walkers are initialized by sampling from the prior distribution. The number of steps it takes for a walker to move from the initial state to the typical set (the region with high posterior mass probability) is known as the burn-in period. 

We herein use a specific MCMC algorithm: the affine invariant MCMC ensembler sample \cite{Goodman2010EnsembleInvariance}, which is implemented in Python through \texttt{emcee} \cite{Foreman-Mackey2013EmceeHammer}. The pivotal idea behind this sampling algorithm is to use multiple chains (referred to as an ensemble) to explore the posterior distribution, and to use information from another chain in the proposal step. By doing so, the sampler is effectively capable of rescaling the parameter domain (referred to as affine invariance), thereby ameliorating sampler performance bottlenecks associated with anisotropic distributions.

\begin{figure}
    \centering    \includegraphics[width=0.5\columnwidth]{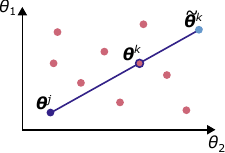}
    \caption{Visualization of a stretch move in the parameter space ($\theta_1$, $\theta_2$). The pink dots represent the current coordinates of each walker. The proposal step for walker $k$, $\tilde{\boldsymbol{\theta}}^{k}$, shown in blue, is based on the current position of walkers $k$ and $j \neq k$.}
    \label{fig:stretch_move}
\end{figure}

We employ the affine invariant MCMC algorithm with its default stretch move, which is visualized in \autoref{fig:stretch_move}. The proposal stretch move is accepted when $r < q$, where $r$ is sampled from a uniform distribution between zero and one and $q$ is given by 
\begin{equation}
    q = \min \left( 1, Z^{p-1} \frac{\pi\left(\tilde{\boldsymbol{\theta}}^{k} \right)}{\pi\left( 
\boldsymbol{\theta}^{k} \right)} \right), 
\end{equation}
where $p$ is the dimension of the parameter space, $\boldsymbol{\theta}^k$ is the current step of walker $k$ and $\tilde{\boldsymbol{\theta}}^{k}$ its proposal step. The proposal is defined by
\begin{equation}
    \tilde{\boldsymbol{\theta}}^{k} =  \boldsymbol{\theta}^{j} +Z \, ( \boldsymbol{\theta}^{k} - \boldsymbol{\theta}^{j}),
\end{equation}
where $\boldsymbol{\theta}^j$ is the current step of a randomly selected walker $j \neq k$ and $Z$ is a random variable sampled from the probability density
\begin{equation}
    g(z) \propto \begin{cases} \frac{1}{\sqrt{z}} & \mathrm{if} \, z \in \left[ \frac{1}{a}, a \right] \\
    0 & \mathrm{otherwise}\end{cases}.
\end{equation}
We herein use the default value of $a=2$.

To investigate whether the chains have sufficiently converged, we apply an autocorrelation analysis by calculating the estimated integrated autocorrelation time $\tau_\theta$ for each parameter $\theta$ in the concatenated chain $\mathscr{C} = \{ \boldsymbol{\theta}_n\}^{N}_{n=1}$. The term ``estimated" comes from the fact that we are looking at finite chains with concatenated length $N$. Subsequent steps in the chain can be dependent on one another, and therefore we need to define the number of independent samples within a chain. The estimated autocorrelation time for a single model parameters is defined as
\begin{equation}
    \tau_{\theta}(M) = 1 + 2\sum_{\tau=1}^M \hat{\rho}_{\theta}(\tau),
\end{equation}
for some $M$\,\(\ll\)\,$N$, where $\hat{\rho}_{\theta}$ is the estimated normalized autocorrelation function of the stochastic process that generated the chain. We use $M$ instead of $N$ to lower the level of noise-to-signal ratio \cite{Sokal1997MonteAlgorithms}. The estimated normalized autocorrelation function is defined as 
\begin{equation}
    \hat{\rho}_{\theta} (\tau) = \frac{\hat{c}_{\theta} (\tau)}{\hat{c}_{\theta} (0)},
\end{equation}
where
\begin{equation}
    \hat{c}_{\theta} (\tau) = \frac{1}{N-\tau}\sum_{n=1}^{N-\tau} (\theta_n - \mu_{\theta})(\theta_{n+\tau} - \mu_{\theta}),
\end{equation}
with
\begin{equation}
    \mu_{\theta} = \frac{1}{N} \sum_{n=1}^{N} \theta_n.
\end{equation}

Based on the maximum autocorrelation time, $\tau = \max_{\theta} \tau_{\theta}$, for each chain we remove the first $N_{\rm burn}=2\cdot \tau$ samples for burn-in, and we thin the chains by selecting every $N_{\rm thin}$-th sample in each chain, with $N_{\rm thin}=\frac{1}{2}\cdot \tau$. This results in a final sample size of 
\begin{equation}
    N_\mathrm{final} = \frac{N - N_\mathrm{walkers} N_\mathrm{burn}}{N_\mathrm{thin}} .
\end{equation}
We note that, preferably, the integrated autocorrelation time is as small as possible to save more of the generated samples instead of discarding them.

\subsection{Rheological characterization}
To characterize the fluids, the simple shear response is measured by performing steady rate sweep tests on a TA Instruments ARES rotational rheometer using a 25\,mm diameter cone-plate geometry. To counteract dehydration of the PVP solution, a Plexiglas ring is attached to the rheometer base around the sample. By injecting silicone oil in between the sample and the Plexiglas ring, the water inside the PVP solution cannot evaporate. We have executed ten measurements for both glycerol and PVP solution. Based on the measurement data and a constitutive model, the likelihood is defined, in accordance with equation~\eqref{eq:likelihood_general}. The average of the experiments is defined as
\begin{equation}
    \boldsymbol{\mu}_{\boldsymbol{y}} = \frac{1}{n} \sum_{i=1}^n \boldsymbol{y}_i, \qquad \boldsymbol{\mu}_{\boldsymbol{y}}, \boldsymbol{y}_i \in \mathbb{R}^{n_t} \quad i=1, ..., n, 
\end{equation}
where $\boldsymbol{\mu}_{\boldsymbol{y}}$ denotes a vector of average values per shear rate and $n$ denotes the number of experiments (ten in this case). The Likelihood equation~\eqref{eq:likelihood_general} can subsequently be written as
\begin{equation}\label{eq:likelihood_sheartime}
L\left(\theta|\boldsymbol{\mu}_{\boldsymbol{y}},\boldsymbol{\Sigma}_{\boldsymbol{y}} \right) \propto \exp \left( -\frac{1}{2} \left( \boldsymbol{d} - \boldsymbol{\mu}_{\boldsymbol{y}} \right)^\mathrm{T} \left[ \boldsymbol{\Sigma}_{\boldsymbol{y}} \right]^{-1} \left( \boldsymbol{d} - \boldsymbol{\mu}_{\boldsymbol{y}} \right) \right),
\end{equation}
where $\boldsymbol{\Sigma}_{\boldsymbol{y}}$ is the covariance matrix corresponding to the measurements.

\subsubsection*{Glycerol}
Based on experience, we assume the constitutive model for glycerol to be Newtonian with a viscosity $\eta_\mathrm{N}$. We define a prior distribution for the viscosity of glycerol. Based on a study performed by Segur \cite{Segur1951ViscositySolutions}, the viscosity of glycerol equals 1.4\,Pa$\cdot$s. Because the viscosity cannot be negative (see \autoref{rmk:logtransform}), we create a prior normal distribution for the log-transform $\hat{\eta}_\mathrm{N}$ with $\mu$=1.4\,Pa$\cdot$s and $\sigma$=0.3\,Pa$\cdot$s, to make the distribution weakly informative. To determine the mean and standard deviation of the lognormal distribution we use
\begin{align}\label{eq:logtransform}
    \hat{\mu} & = \ln \left( \frac{\mu^2}{\sqrt{\mu^2 + \sigma^2}} \right), \qquad 
    \hat{\sigma}^2 = \ln \left( 1 + \frac{\sigma^2}{\mu^2} \right).
\end{align}

To sample from the posterior, we define the number of walkers/chains, number of samples per walker, the burn-in period and the thinning. We have used four walkers and 10,000 samples per walker. The burn-in period is defined as $2\times\tau_f$ and the thinning as $\tau_f$/2, where the autocorrelation time $\tau_f$ for $\hat{\eta}_\mathrm{N}$ equals 30. The effective sample size $N_\mathrm{final}$ then equals 3,059 samples. We evaluated $N=4\times10,000=40,000$ samples out of which 3,059 have been used to make a model prediction, indicating that a substantial number of samples has been lost due to the burn-in period and (especially) thinning.  

\begin{figure}
    \centering
    \includegraphics[width=\columnwidth]{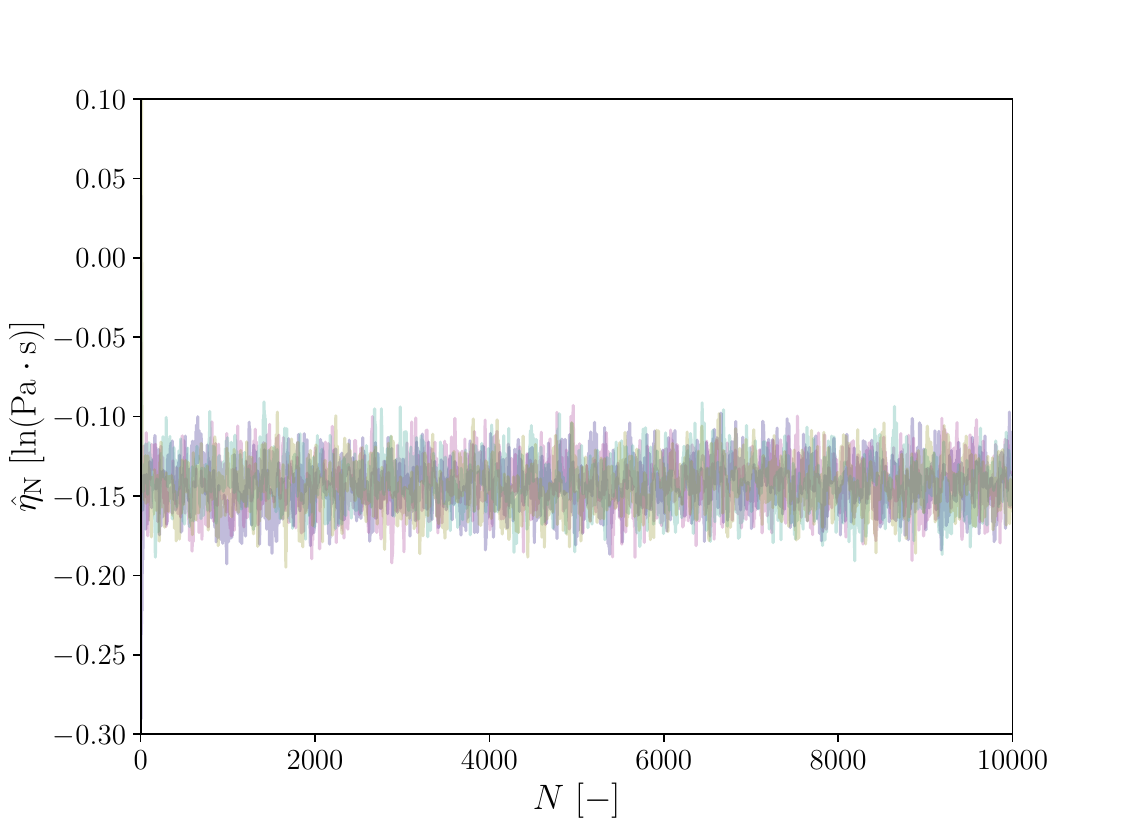}
    \caption{Trace plot of of the log-transformed Newtonian viscosity $\hat{\eta}_\mathrm{N}$ using four walkers and 10,000 samples.}
    \label{fig:tracerheogly}
\end{figure}

We construct a trace plot to investigate whether sufficient samples have been used to define the posterior distribution. If the chains fluctuate around a value for $\hat{\eta}_\mathrm{N}$, we can assume that sufficient samples have been generated for the chain to be stationary. The trace plot for $\hat{\eta}$ is shown in \autoref{fig:tracerheogly}. We observe that all the walkers fluctuate around a horizontal line and therefore a certain value for $\hat{\eta}_\mathrm{N}$.

\begin{figure}
     \centering
     \includegraphics[width=\columnwidth]{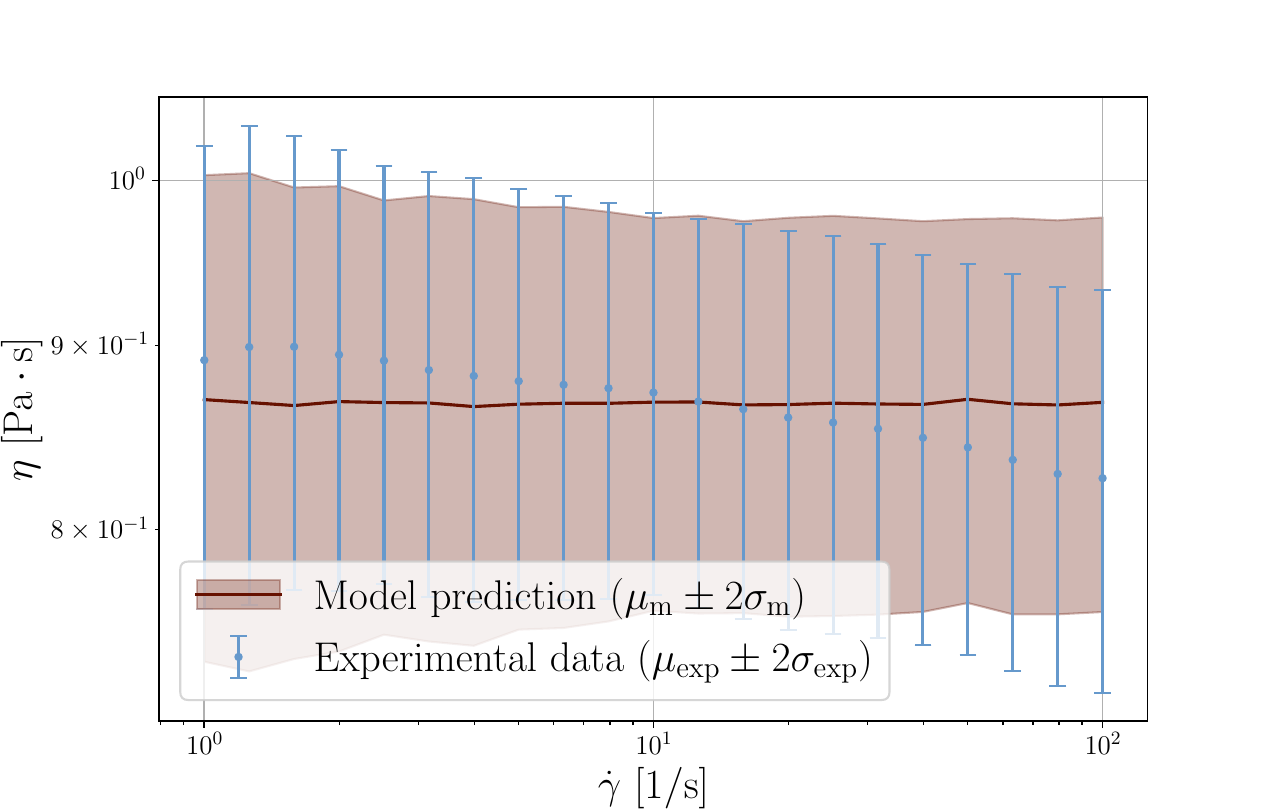} 
    \caption{Posterior predictive distribution of $\eta_\mathrm{N}$ for glycerol using a Newtonian constitutive model, where the subscript 'm' corresponds to the model and the subscript 'exp' to the experimental data. Because of the logarithmic scale, the error bars are asymmetric.}
    \label{rheodataGLY}  
\end{figure} 

The posterior predictive distribution of the rheological model is shown in \autoref{rheodataGLY}, together with the experimental data. Note that, as discussed in Section~\ref{sec:uq}, experimental observations include both parametric uncertainties and observation errors. Including the latter in the model predictions yields the posterior predictive distribution (PPD) \cite{Rappel2020AMechanics}. We determine the PPD by sampling from the posterior (reusing samples from the MCMC), determining the model response and adding noise according to the noise model \eqref{eq:likelihood_sheartime}. Alternatively, the uncertainty in the model results could have been visualized through the posterior distribution, which yields probabilistic model realizations that only include parametric uncertainties.

\begin{figure}
     \centering
     \subfloat[]{\includegraphics[width=0.49\columnwidth]{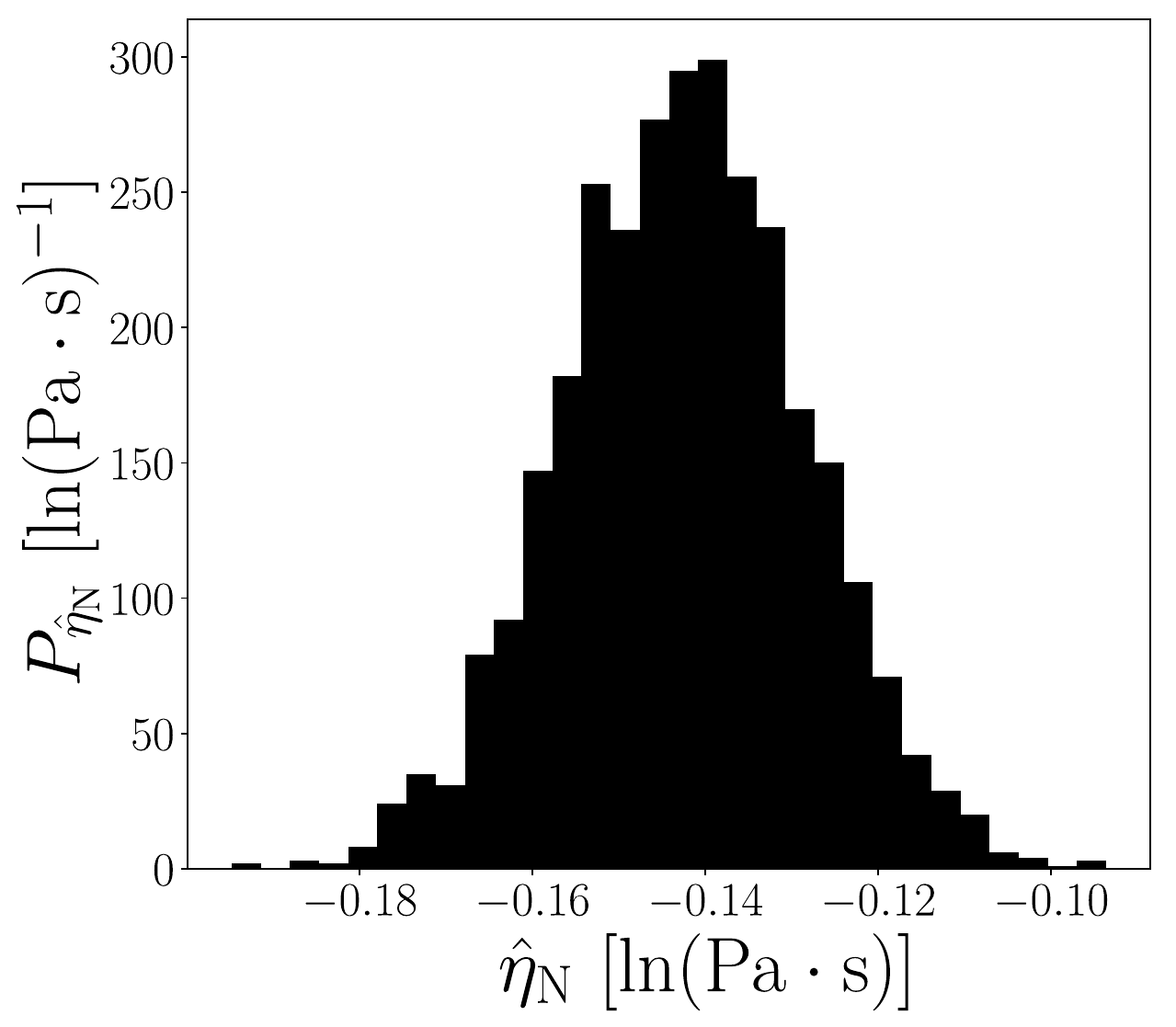}}\label{fig:hist_etaNdakje}
     \hfill
     \subfloat[]{\includegraphics[width=0.49\columnwidth]{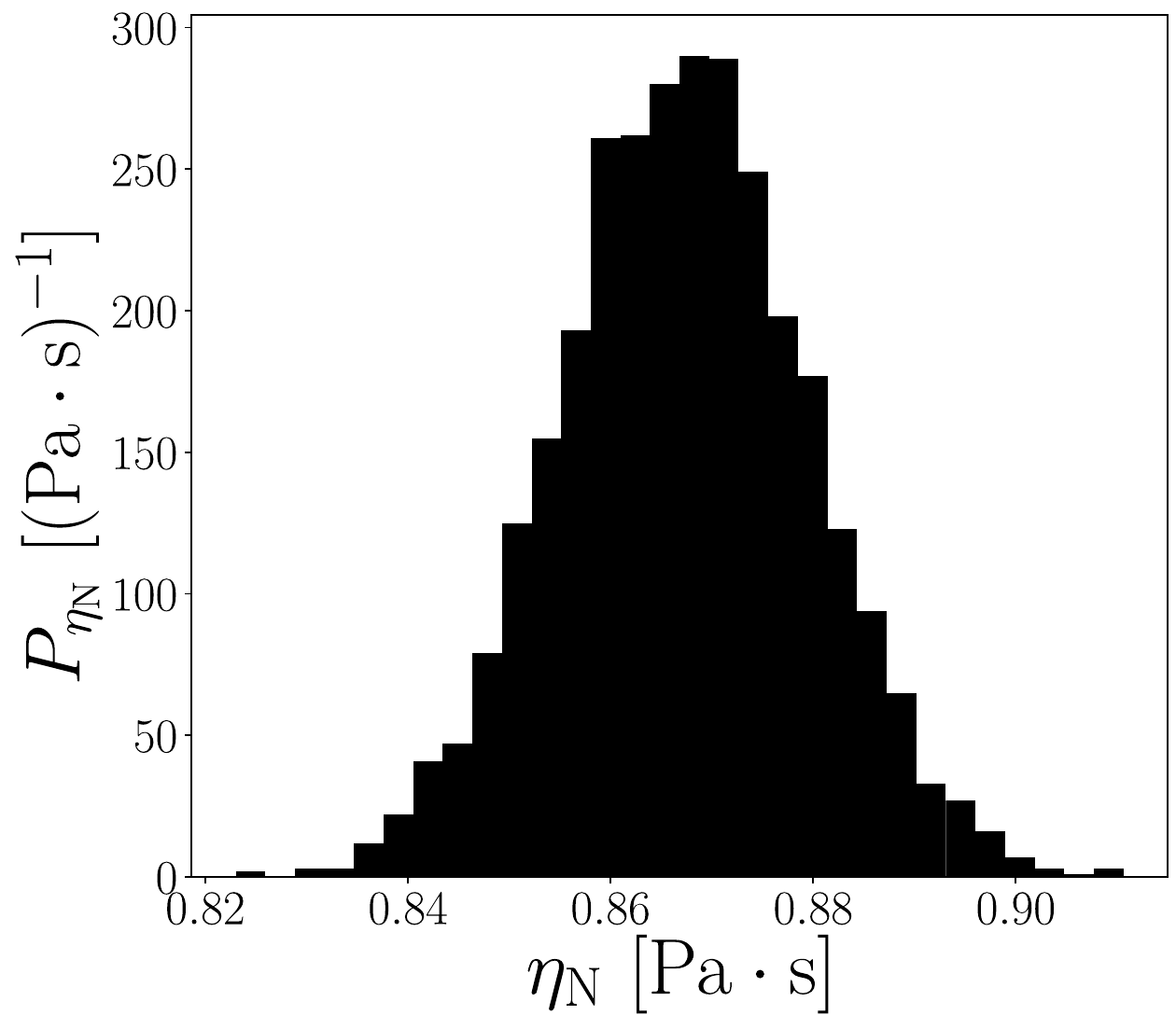}\label{fig:hist_etaN}}
     \caption{Comparison between the histogram of the model parameter $\eta_\mathrm{N}$ and its log-transform $\hat{\eta}_\mathrm{N}=\ln\left( \eta_\mathrm{N} \right)$.}
     \label{fig:logtransformetaN}
\end{figure}

Because the model is equal to the constant value $\hat{\eta}_\mathrm{N}$, the viscosity is independent of the shear rate and equal to the posterior distribution $\hat{\eta}_\mathrm{N}$. The posterior distribution for $\eta_\mathrm{N}$ is visualized in \autoref{fig:logtransformetaN} together with the posterior distribution $\hat{\eta}_\mathrm{N}$. The posterior distribution for $\eta_\mathrm{N}$ is obtained by calculating the exponential function of $\hat{\eta}_\mathrm{N}$ ($\eta_\mathrm{N}=\exp\left(\hat{\eta}_\mathrm{N}\right)$) for each sample and creating a distribution from these samples. We observe that both distributions have a similar shape, but have a different mean and standard deviation. Because the standard deviation of $\eta_\mathrm{N}$ is relatively small and not close to zero, the right-skewness characteristic for a log-transform distribution is not clearly visible. To evaluate the level of uncertainty in $\eta_\mathrm{N}$, the coefficient of variation, $\mathrm{CV} = (\sigma / \mu) \times 100$,  is determined. The mean $\mu$, standard deviation $\sigma$ and coefficient of variation CV are provided in \autoref{tab:PU_rheoGly}.

\begin{table}
    \centering
    \caption{Uncertainty in the rheological parameter of glycerol.}
    \begin{tabularx}{\columnwidth}{XXXX}
    \toprule
         & $\mu$ [Pa$\cdot$s] &  $\sigma$ [Pa$\cdot$s] & CV [\%]\\ \midrule
        $\eta_\mathrm{N}$ & 8.67$\,\times\,10^{-1}$ & 1.22$\,\times\,10^{-2}$ & 1.41 \\ \bottomrule
    \end{tabularx}
    \label{tab:PU_rheoGly}
\end{table}

\subsubsection*{PVP solution}
The truncated power law model parameters $\eta_0$, $\eta_\infty$ and $\lambda_\text{cr}$, which can physically not be negative, are approached in a similar way as the viscosity $\eta_\mathrm{N}$ in the Newtonian model. The normal distributions of the log-transforms of these parameters are denoted by $\hat{\eta}_0$, $\hat{\eta}_\infty$ and $\hat{\lambda}_\text{cr}$. We assign a uniform distribution between zero and one to $n$, because the material shows shear thinning behavior and $n<0$ would lead to a non-physical maximum in the shear stress. 

We have used eight walkers and 20,000 samples per walker. The burn-in period and thinning are defined similarly as for the inference applied on the Newtonian model for glycerol. The autocorrelation times for the rheological model parameters are provided in \autoref{tab:autocor_rheoPVP}. The total number of samples $N$, and thus the number of physical model evaluations, equals 160,000. The final sample size is equal to 12,239, which is based on the highest autocorrelation time of the model parameters.
\begin{table}
    \centering
    \caption{Autocorrelation times per model parameter for PVP solution using eight walkers and 20,000 samples per walker.}
    \begin{tabularx}{\columnwidth}{XXXX}
    \toprule
         $\hat{\lambda}_\mathrm{cr}$ & $\hat{\eta}_0$ & $\hat{n}$ & $\hat{\eta}_\infty$ \\ \midrule
        48 & 74 & 66 & 44\\ \bottomrule 
    \end{tabularx}
    \label{tab:autocor_rheoPVP}
\end{table}

\begin{figure}
     \centering
     \includegraphics[width=\columnwidth]{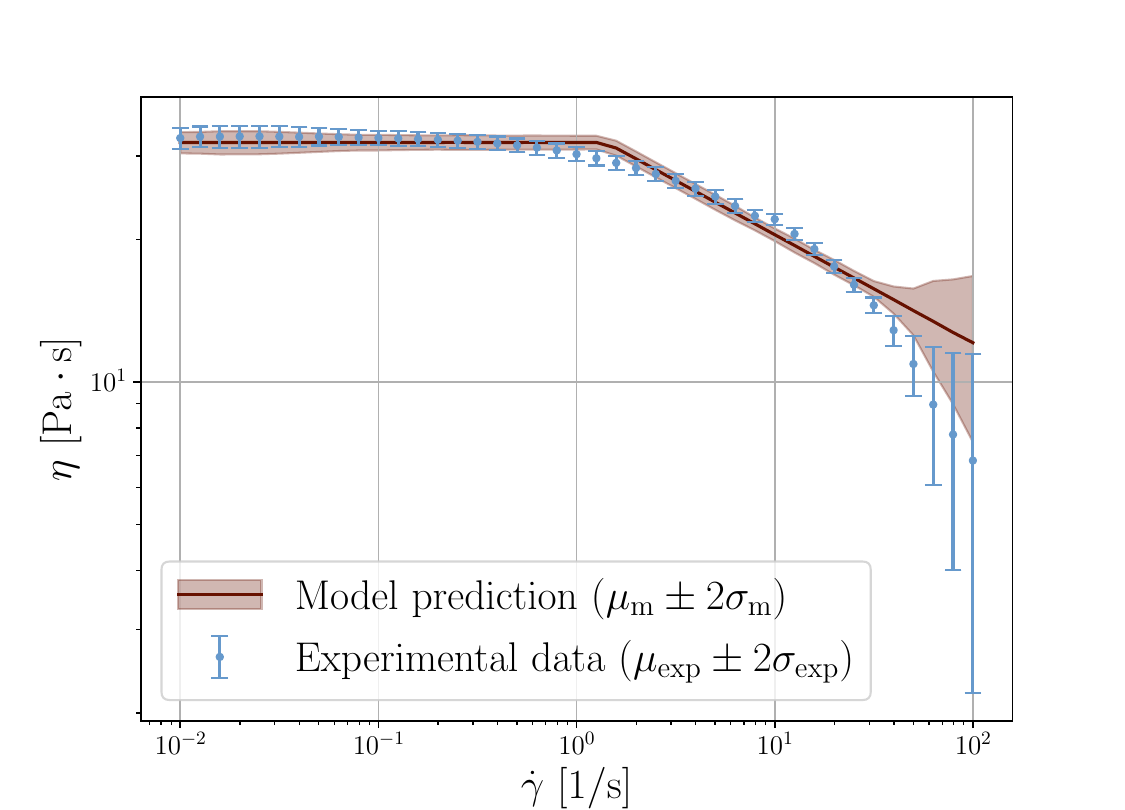}  
    \caption{Posterior predictive distribution of $\eta_{\mathrm{TPL}}(\dot{\gamma})$ using the truncated power law constitutive model, where the subscript 'm' corresponds to the model and the subscript 'exp' to the experimental data. Because of the logarithmic scale, the error bars are asymmetric per shear rate and appear to increase as the shear rate increases.}
    \label{fig:rheodataPVP}  
\end{figure}

\begin{figure}
    \centering    \includegraphics[width=\columnwidth]{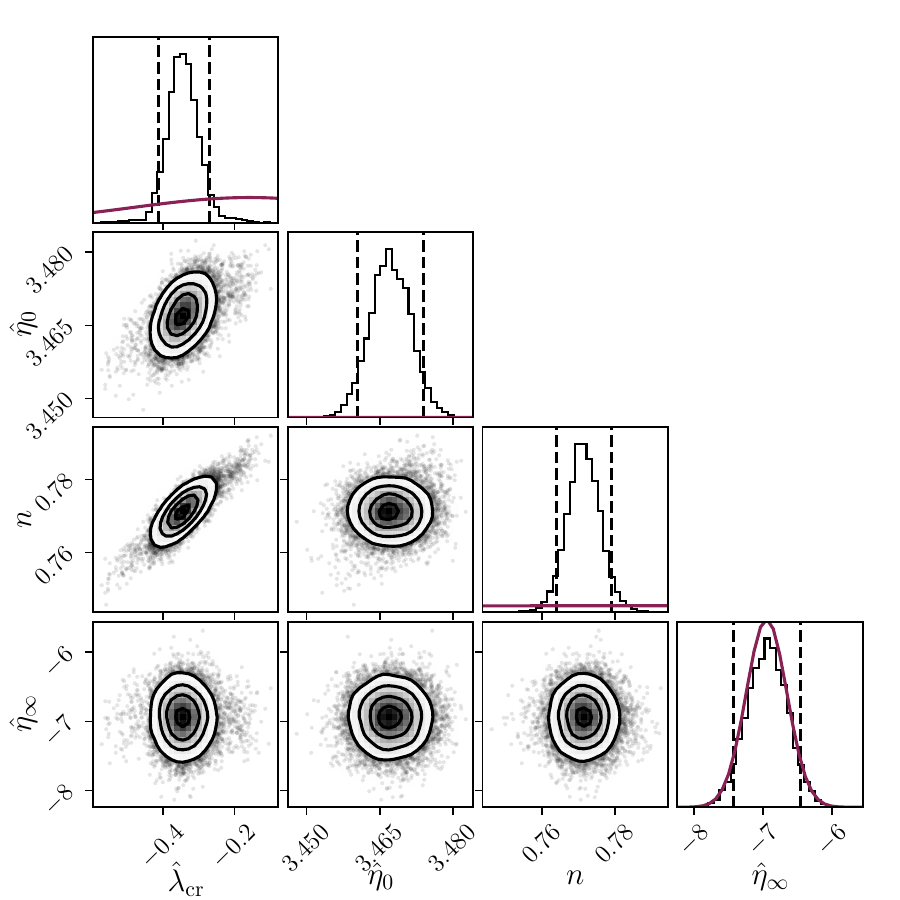}
    \caption{Corner plots showing the correlation between the truncated power law parameters. The histogram and dark-red graphs denote the posterior and prior, respectively. The region in between the black dashed lines defines the 95\% credibility interval.}
    \label{fig:CornerrheoPVP}
\end{figure}

The constitutive model prediction and experimental results are shown in \autoref{fig:rheodataPVP}.
The distribution of and correlation between the model parameters $\hat{\eta}_0$, $\hat{\eta}_\infty$, $\hat{\lambda}_\text{cr}$ and $n$ are shown in \autoref{fig:CornerrheoPVP} \cite{Foreman-Mackey2016Corner.py:Python}. The prior distributions are denoted by the solid red line and the histogram denotes the posterior distribution for the model parameters. The dotted black lines represent the 95\% ``credibility interval”, \emph{i.e.}, the interval containing 95\% of the area under the posterior distribution. The samples and distribution of the correlation between every two parameters are provided in the two-dimensional posterior marginals. We find that $n$ and $\hat{\eta}_0$ show a strong dependence on the likelihood because the prior probability is low and flat as compared to the posterior, while $\hat{\eta}_\infty$ is mainly determined by the prior information. The influence of the likelihood and prior are both visible in the posterior distribution of $\hat{\lambda}_\text{cr}$. The correlation plots indicate that the strongest correlation exists between $n$ and $\hat{\eta}_0$. The statistical moments are provided in \autoref{tab:PU_rheoPVP}.

\begin{table}
    \centering
    \caption{Uncertainty in rheological parameters of PVP solution.}
    \begin{tabularx}{\columnwidth}{XXXX}
    \toprule
         & $\mu$ &  $\sigma$ & CV [\%]\\ \midrule
        $\lambda_\mathrm{cr}$ [1/s] & 7.12$\,\times\,10^{-1}$ & 3.33$\,\times\,10^{-2}$ & 4.69 \\
        $\eta_0$ [pa$\cdot$s] & 3.20$\,\times\,10^{1}$ & 1.34$\,\times\,10^{-1}$ & 4.19$\,\times\,10^{-1}$ \\
       $n$ [-]& 7.71$\,\times\,10^{-1}$ & 4.72$\,\times\,10^{-3}$ & 6.11$\,\times\,10^{-1}$ \\ 
       $\eta_\infty$ [pa$\cdot$s] & 1.01$\,\times\,10^{-3}$ & 3.05$\,\times\,10^{-4}$ & 3.02$\,\times\,10^{1}$ \\ \bottomrule
    \end{tabularx}
    \label{tab:PU_rheoPVP}
\end{table}

\subsection{Independent parameter experiments}\label{PUfluidvolume}
Because of the different cases used for the squeeze flow experiments, several distributions are obtained for the fluid volume, applied force and initial radius. Each of the parameters is given a (log-)normal distribution where the mean and standard deviation are based on ten measurements. 
\begin{itemize}
    \item \textbf{Fluid volume}. The uncertainty in fluid volume is determined differently for glycerol and PVP solution. For glycerol, the fluid volume is measured separately from the squeeze flow experiment. We pipette ten samples and weigh them to find the distribution in fluid volume, using the density of glycerol as obtained from the supplier. For PVP solution, each sample is weighted just before performing the squeeze flow experiment. The density of PVP solution is experimentally determined by weighing a sample of known volume. The fluid volume distribution applicable to several squeeze flow experimental cases are given in \autoref{tab:PU_fluidvolume}.

\begin{table}
    \centering
    \caption{Uncertainty in fluid volume used for the squeeze flow cases. $V_a$ is used for cases \RNum{1}, \RNum{2} and \RNum{3}. $V_b$ is used for case \RNum{4} (see also \autoref{tab:expcases}).}
    \begin{tabularx}{\columnwidth}{XXXX}
    \toprule
         & $\mu$ [mL]&  $\sigma$ [mL] & CV [\%]\\ \midrule
        $V_a$ & 2.04$\,\times\,10^{-1}$ & 2.22$\,\times\,10^{-2}$ & 10.9 \\
        $V_b$ & 2.96$\,\times\,10^{-1}$ & 1.38$\,\times\,10^{-2}$ & 4.67 \\
       $V_{\RNum{5}}$ & 5.53$\,\times\,10^{-1}$ & 6.62$\,\times\,10^{-2}$ & 12.0 \\ 
       $V_{\RNum{6}}$ & 5.75$\,\times\,10^{-1}$ & 5.28$\,\times\,10^{-2}$ & 9.17 \\ 
       $V_{\RNum{7}}$ & 1.29$\,\times\,10^{-1}$ & 1.16$\,\times\,10^{-2}$ & 8.94 \\ 
       $V_{\RNum{8}}$ & 1.59$\,\times\,10^{-1}$ & 2.24$\,\times\,10^{-2}$ & 14.1 \\ \bottomrule
    \end{tabularx}
    \label{tab:PU_fluidvolume}
\end{table} 

    \item \textbf{Applied force}. We measure the applied force using a spring suspension attached to the moving parallel plates structure, which is denoted by the black part in \autoref{fig:setup_schem}. We perform force-displacement measurements for several combinations of force and displacement to obtain the spring constant of the construction. Due to the minimal vertical displacement, the force is assumed to be constant in this range. The mean and standard deviation are determined from the linear fit onto the measured data. The applied force distributions are provided in \autoref{tab:PU_Fapp}. 
    \begin{table}
        \centering
        \caption{Uncertainty in applied force for three different additions of weight. $F_{\mathrm{add},1}$, $F_{\mathrm{add},2}$ and $F_{\mathrm{add},3}$ correspond to 0.0 kg, 0.25 kg and 0.5 kg, respectively.}
        \begin{tabularx}{\columnwidth}{XXXX}
        \toprule
             & $\mu$ [N] &  $\sigma$ [N] & CV [\%]\\ \midrule
           $F_{\mathrm{add},1}$ & 2.89 & 6.04$\,\times\,10^{-2}$ & 2.09  \\ 
           
           $F_{\mathrm{add},2}$ & 5.79 & 6.50$\,\times\,10^{-2}$ & 1.12\\ 
           
           $F_{\mathrm{add},3}$ & 8.56 & 6.52$\,\times\,10^{-2}$ & 7.61$\,\times\,10^{-1}$\\ \bottomrule
        \end{tabularx}
        \label{tab:PU_Fapp}
    \end{table}

    \item \textbf{Initial radius}. The initial radius is obtained from the image processing step by averaging the values of initial radius obtained from a single set of the squeeze flow measurements. Each squeeze flow case corresponds to a different distribution of the initial radius. The initial radius distributions are provided in \autoref{tab:PU_initialradius}. 
    
    \begin{table}
        \centering
        \caption{Uncertainty in initial radius for the squeeze flow cases.}
        \begin{tabularx}{\columnwidth}{XXXX}
        \toprule
             & $\mu$ [cm] &  $\sigma$ [cm] & CV [\%]\\ \midrule
           $R_{0,\RNum{1}}$ & 7.55$\,\times\,10^{-3}$ & 1.50$\,\times\,10^{-4}$ & 1.98 \\ 
           $R_{0,\RNum{2}}$ & 7.12$\,\times\,10^{-3}$ & 2.39$\,\times\,10^{-4}$ & 3.36 \\ 
           $R_{0,\RNum{3}}$ & 7.33$\,\times\,10^{-3}$ & 2.29$\,\times\,10^{-4}$ & 3.12 \\ 
           $R_{0,\RNum{4}}$ & 8.21$\,\times\,10^{-2}$ & 2.85$\,\times\,10^{-4}$ & 3.47 \\ 
           $R_{0,\RNum{5}}$ & 1.18 & 6.30$\,\times\,10^{-2}$ & 5.33 \\ 
           $R_{0,\RNum{6}}$ & 1.22 & 5.74$\,\times\,10^{-2}$ & 4.70 \\ 
           $R_{0,\RNum{7}}$ & 5.62$\,\times\,10^{-1}$ & 2.14$\,\times\,10^{-2}$ & 3.80 \\ 
           $R_{0,\RNum{8}}$ & 6.38$\,\times\,10^{-1}$ & 4.66$\,\times\,10^{-2}$ & 4.30 \\ \bottomrule
        \end{tabularx}
        \label{tab:PU_initialradius}
    \end{table}

    \item \textbf{Laplace pressure}. To account for the surface tension in the squeeze flow model, we implement the Laplace pressure (equation~\eqref{eq:Laplace}), which depends on the surface tension $\gamma$ and curvature of the interface $\kappa$. The uncertainty in surface tension of glycerol has been measured by pendant drop experiments, of which the mean, standard deviation and coefficient of variation are provided in \autoref{tab:PU_gamma}. We have quantified the surface tension for PVP solution using literature \cite{Bolten2011ExperimentalSolutions}. 
    \begin{table}
        \centering
        \caption{Uncertainty in surface tension.}
        \begin{tabularx}{\columnwidth}{XXXX}
        \toprule
             & $\mu$ [N/m] &  $\sigma$ [N/m] & CV [\%]\\ \midrule
           $\gamma$ glycerol& 4.54$\,\times\,10^{-2}$ & 2.53$\,\times\,10^{-3}$ & 5.57  \\ 
           $\gamma$ PVP solution & 6.60$\,\times\,10^{-2}$ & 2.00$\,\times\,10^{-3}$ & 3.03 \\\bottomrule
        \end{tabularx}
        \label{tab:PU_gamma}
    \end{table}
    
    Quantifying the curvature at the fluid front is more challenging, as it is not directly observable in the squeeze flow experiment. The curvature is implemented with a dependence on the semi-height of the flow domain $h$ as
    \begin{equation}\label{eq:curvekappa}
        \kappa = \frac{\alpha}{h},
    \end{equation}
    where $\alpha$ is a factor valued between 0 and 1. For $\alpha=0$, the interface 
    is straight, and thus the Laplace pressure is excluded. For $\alpha=1$, we have reached the maximum curvature, \emph{i.e.},  $\kappa=1/h$.
\end{itemize}

\section{Comparison of the models and experiments}
\label{sec:results}

In this section we present the comparison between the experimental and numerical results for the squeeze flow of glycerol and PVP solution. The comparison is realized by uncertainty propagation, using the forward Monte Carlo sampling method, and Bayesian inference, using MCMC. 

We used Bayesian inference on the constitutive model to obtain a distribution for the rheological parameters (see Section~\ref{sec:uncertainties}). Based on the simple shear measurements and constitutive model prediction, we expect that the squeeze flow behavior of glycerol can be described by the Newtonian squeeze flow model. For PVP, a prediction for the initial shear rate using the Newtonian model and zero-shear viscosity, yields maximum shear rates in the order of 10$^2$\,1/s. Since this is in the shear-thinning regime (\autoref{fig:rheodataPVP}), we use the truncated power law model to predict the squeeze flow behavior of PVP solution.

For both of the fluids, we start by using the method of uncertainty propagation to compare the numerical model to the experimental data, keeping the parameter distributions fixed. Thereafter, we use Bayesian inference to update the parameters, and thus the model response, using the experimental data from the squeeze flow. Finally, we will compare both methods for both fluids. 

\subsection{Newtonian fluid}
The results of the experimental cases \RNum{1} to \RNum{4} are visualized in \autoref{fig:UP_Gly_NLP01}. From case \RNum{1} to \RNum{3} we have an increase in applied force (see \autoref{tab:ExpCaseGly} and \autoref{tab:PU_Fapp}). In \autoref{fig:UP_Gly_NLP01} we observe that an increase in applied force leads to a faster growth of the radius. The applied force in case \RNum{4} is similar to the applied force in case \RNum{2} and the volume is increased (see \autoref{tab:ExpCaseGly} and \autoref{tab:PU_fluidvolume}). We observe that an increase in fluid volume leads to a slight increase in initial radius. Analyzing the experimental data, we see that the uncertainty increases as time proceeds.

\subsubsection*{Uncertainty propagation}\label{sec:UP_gly}
To use the method of uncertainty propagation, we assign a distribution to the model input parameters. In Section~\ref{sec:uncertainties}, we have discussed experiments to obtain the uncertainty in the input parameters $F$, $V$, $R_0$ and $\gamma$ and have obtained the distributions for these parameters. The curvature $\alpha$ has been defined in two limiting cases: $\alpha=0$ and $\alpha=1$. Using Bayesian inference on the simple shear measurements, we have retrieved a distribution for the rheological parameters. Every distribution is to be propagated through the squeeze flow model to obtain the distribution of the radius over time. We have used 10,000 samples from each distribution of the model parameters to retrieve 10,000 samples for the radius per point in time.   

\begin{figure}
    \centering 
    \includegraphics[width=\columnwidth, trim={0.5cm 1.25cm 1.25cm 0}, clip]{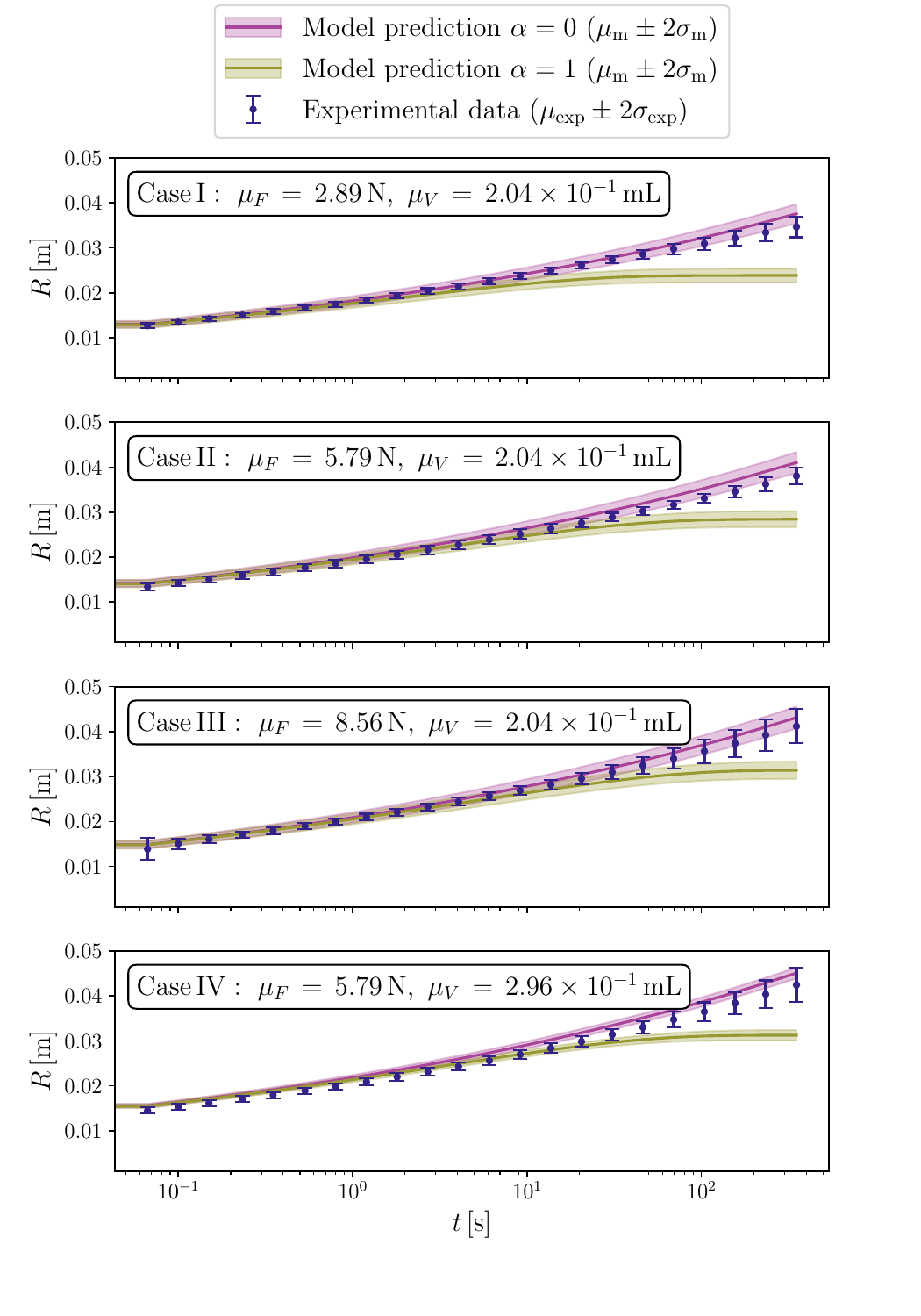}
    \caption{Comparison between the experimental data of glycerol and the model prediction using uncertainty propagation. The subscript `m' denotes the model prediction and the subscript `exp' denotes the experimental data. The model prediction including the Laplace pressure is the case where we have the maximum possible curvature. The two extreme cases encompass the experimental data for all cases (\RNum{1} to \RNum{4}).}
    \label{fig:UP_Gly_NLP01}
\end{figure}

We consider the Newtonian model prediction including the two limiting cases of the Laplace pressure ($\alpha=0$ and $\alpha=1$) in \autoref{fig:UP_Gly_NLP01}. In the case of $\alpha=0$ (no Laplace pressure), we observe that the model prediction increasingly deviates from the experimental data as time proceeds. 
Moreover, the difference between the model prediction without Laplace pressure and the experimental data decreases with increased applied force. These observations can be explained by the capillary number $\mathrm{Ca}=\eta \dot{h} / \gamma$. In the squeeze flow setting, the radius increases with time meaning that $\dot{h}$ will decrease and thus also Ca. The capillary number decreases from $\sim$ $10^{-2}$ at $t=0$\,s to $\sim$ $10^{-7}$ at around $t=350$\,s in case \RNum{1}. 
A low capillary number indicates that the influence of the surface tension increases. This explains why the model prediction without Laplace pressure describes the short-term flow behavior well in comparison to the long-term flow behavior. It also explains why there is an increasing similarity between the experimental data and model prediction for an increasing applied force. 

In \autoref{fig:UP_Gly_NLP01}, we can see that for $\alpha=1$ (maximum Laplace pressure) for cases \RNum{1} to \RNum{4}, denoted in yellow, the model response clearly underestimates the experimental data. Based on the observation in \autoref{fig:UP_Gly_NLP01}, the curvature should be smaller than the maximum curvature to describe the experimental data.
Because we are uncertain about the value of $\alpha$ throughout the experiment, we have a relatively wide range of possible model predictions. In order to quantify the curvature we use the method of Bayesian inference.  

\subsubsection*{Bayesian inference}
In the framework of Bayesian inference, we use the experimental squeeze flow data to define the likelihood and use the parametric uncertainties as priors. The likelihood function is defined in a similarly way as the one used in Bayesian inference on the rheological measurements (see Section~\ref{sec:uncertainties}), where we use the experimental data to determine the noise in the likelihood function. However, in this case the noise differs per time step instead of per shear rate.The procedure of Bayesian inference is conducted using the MCMC method described in Section~\ref{sec:uncertainties}. 

The prior distribution of the model input parameters is similar to the distribution used in the method of uncertainty propagation. For these parameters, we choose normal distributions. The parameters $\eta$, $R_0$ and $V$ should physically be non-negative. Therefore, we use equation~\eqref{eq:logtransform} to obtain a lognormal distribution. We use the transformation explained in \autoref{rmk:logtransform} to obtain the distribution of the true parameter values.
We define $\alpha$ in between zero and one. Based on the observations made in \autoref{fig:UP_Gly_NLP01}, we expect that $\alpha$ is closer to zero than to one. Therefore, we assign a beta distribution to $\alpha$, of which the probability density function is defined as
\begin{equation}\label{eq:betadist}
    f(\alpha;a,b) = \frac{1}{B(a,b)} \alpha^{a-1} (1-\alpha)^{b-1} \qquad \alpha \in [0,1],
\end{equation}
where $a$ and $b$ are the shape parameters valued $a=3$ and $b=8$, respectively and $B(a,b)$ is a normalization constant.

We evaluate the Newtonian squeeze flow model using twelve walkers/chains and 10,000 samples per walker. The burn-in period per walker is $2\times\tau_f$ and the thinning is $\tau_f$/2, similar to the formulation as described in Section~\ref{sec:uncertainties}. The autocorrelation time per model parameter for each of the cases \RNum{1} to \RNum{4} is given in \autoref{tab:autocor_Gly}. The effective sample size per case is given in \autoref{tab:Neff_Gly}.

\begin{table}
    \centering
    \caption{Autocorrelation times per model parameter and case using twelve walkers and 10,000 samples per walker.}
    \begin{tabularx}{\columnwidth}{XXXXXXX}
    \toprule
         Case &  $F$ & $\hat{V}$ & $\hat{R}_0$ & $\hat{\eta_\mathrm{N}}$ & $\gamma$ & $\alpha$ \\ \midrule
        \RNum{1} & 79 & 76 & 87 & 82 & 83 & 83 \\ 
        \RNum{2} & 96 & 128 & 158 & 110 & 84 & 127 \\
        \RNum{3} & 103 & 92 & 84 & 72 & 96 & 95 \\
        \RNum{4} & 108 & 82 & 102 & 72 & 120 & 82 \\ \bottomrule
    \end{tabularx}
    \label{tab:autocor_Gly}
\end{table}
\begin{table}
    \centering
    \caption{Effective sample size per case based on the highest autocorrelation time per case for a sample size $N=120,000$.}
    \begin{tabularx}{\columnwidth}{XXXXX}
    \toprule
        Case & \RNum{1} & \RNum{2} & \RNum{3} & \RNum{4} \\ \midrule
        $N_\mathrm{final}$ & 2736 & 1488 & 2304 & 1944 \\ \bottomrule
    \end{tabularx}
    \label{tab:Neff_Gly}
\end{table}

\begin{figure}
    \centering
    \includegraphics[width=\columnwidth, trim={0.5cm 1cm 0.5cm 0cm}, clip]{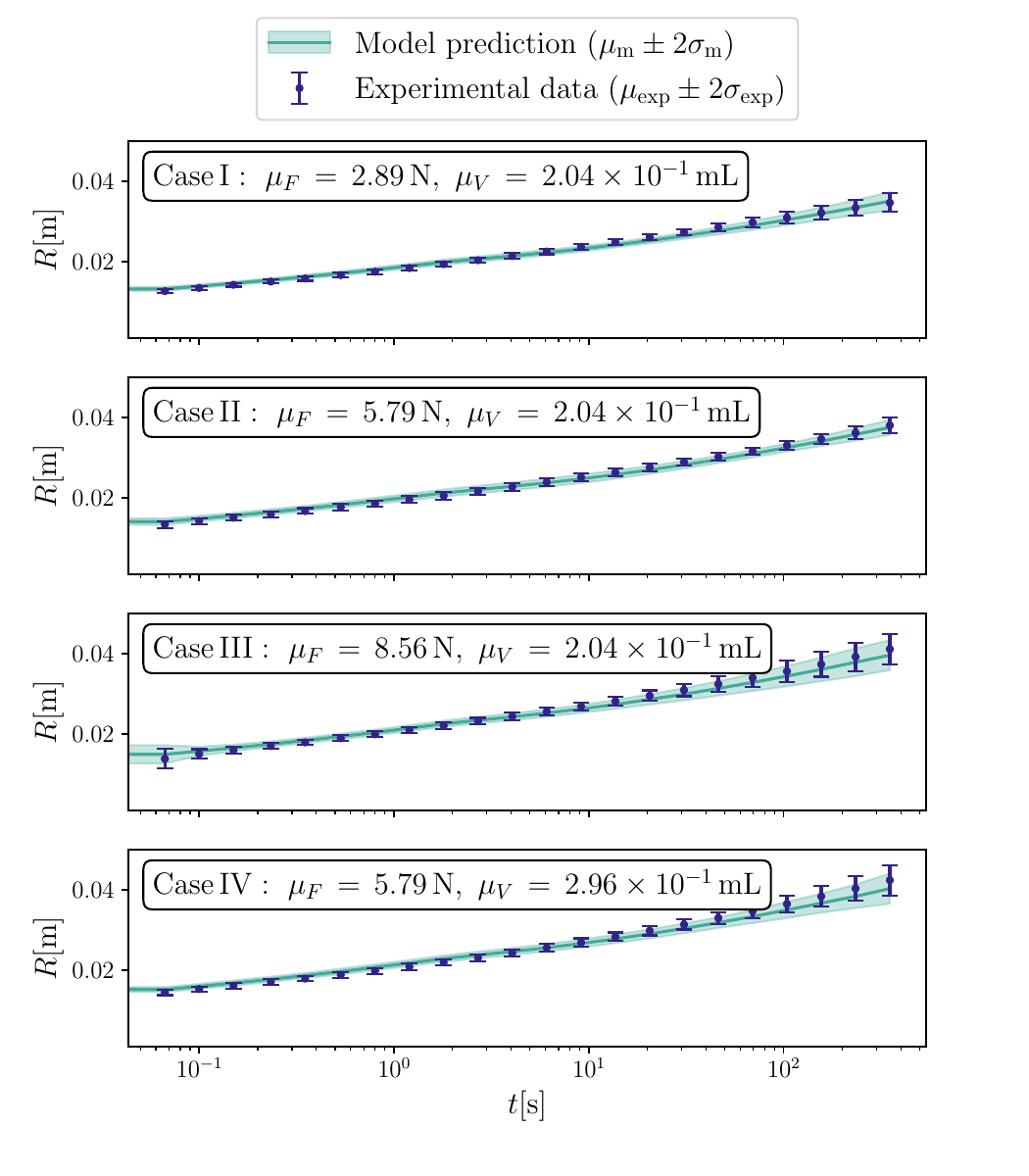}
    \caption{Comparison between the posterior predictive distribution and experimental data of the squeeze flow using Bayesian inference. The subscript 'm' corresponds to the model and the subscript 'exp' to the experimental data.}
    \label{fig:BI_Gly_NLP_ppd}
\end{figure}

In \autoref{fig:BI_Gly_NLP_ppd}, we show the posterior predictive distribution of the model response $R(t)$ and the experimental data for cases \RNum{1} to \RNum{4}. The width of the model prediction is comparable to the width of the experimental data. The initial data point in case \RNum{3} shows an increased distribution width for the posterior predictive distribution as well as the experimental data, because the noise in the posterior predictive distribution is determined using the standard deviation of the experimental data, which differs per time step. The short-term flow behavior predicted using uncertainty propagation shows similar behavior to the prediction made using Bayesian inference, because the Laplace pressure's influence decreases toward the short-term flow behavior. Furthermore, the similarities in model response determined using uncertainty propagation and Bayesian inference indicates that the uncertainty in the model input parameters are well-quantified, meaning that the parametric uncertainty determined using independent parameter experiments shows a similar amount of uncertainty as the parametric uncertainty obtained using the squeeze flow experimental data.

The posterior of the model parameters and the correlation between them are visualized in \autoref{fig:BI_Gly_NLP_corner} for case \RNum{1}. Similar plots for the remaining cases can be found in the supplementary information. The posterior per input parameter is visualized by the histogram and the prior distribution by the green graph. Furthermore, the area in between the dashed black lines indicates the 95\% credibility interval. We observe that for the parameters $\eta_\mathrm{N}$, $F$ and $\gamma$ the posterior looks similar to the prior distribution, indicating that the model parameters are not updated through the likelihood function. We also observe that the prior of the fluid volume is not similar to the posterior. This suggests that the pipetting and subsequent weighting of the sample yield predicted volumes that
are inconsistent with the squeeze flow model predictions. The posterior of $\alpha$ is not similar to the prior distribution and is close to zero, suggesting that the influence of the Laplace pressure is minimally present. We would expect a perfect correlation between $\alpha$ and $\gamma$ based on the expression for the Laplace pressure (see equation~\eqref{eq:Laplace} and equation~\eqref{eq:curvekappa}). However, because we are much more certain about $\gamma$ in comparison to $\alpha$, the uncertainty is dominated by $\gamma$, due to which the parameters do not seem to be correlated. The posteriors in the corner plot are given as $\hat{\boldsymbol{\theta}}$ for the parameters to which we have applied the log-transformation. The mean, standard deviation and coefficient of variation are given in \autoref{tab:musig_BI_squeezeGly} for the actual model parameters $\boldsymbol{\theta}$. For the model parameters that are difficult to determine, such as the curvature-related parameter $\alpha$, Bayesian inference can give an accurate probablistic prediction. The uncertainty in viscosity and surface tension is expected to be similar for cases \RNum{1} to \RNum{4}. In \ref{app:uncertaintytab} we show the uncertainty for the remaining cases (cases \RNum{2} to \RNum{4}), from which it can be concluded that the uncertainty per case is comparable. 

\begin{figure*}
    \centering
    \includegraphics[width=0.7\textwidth]{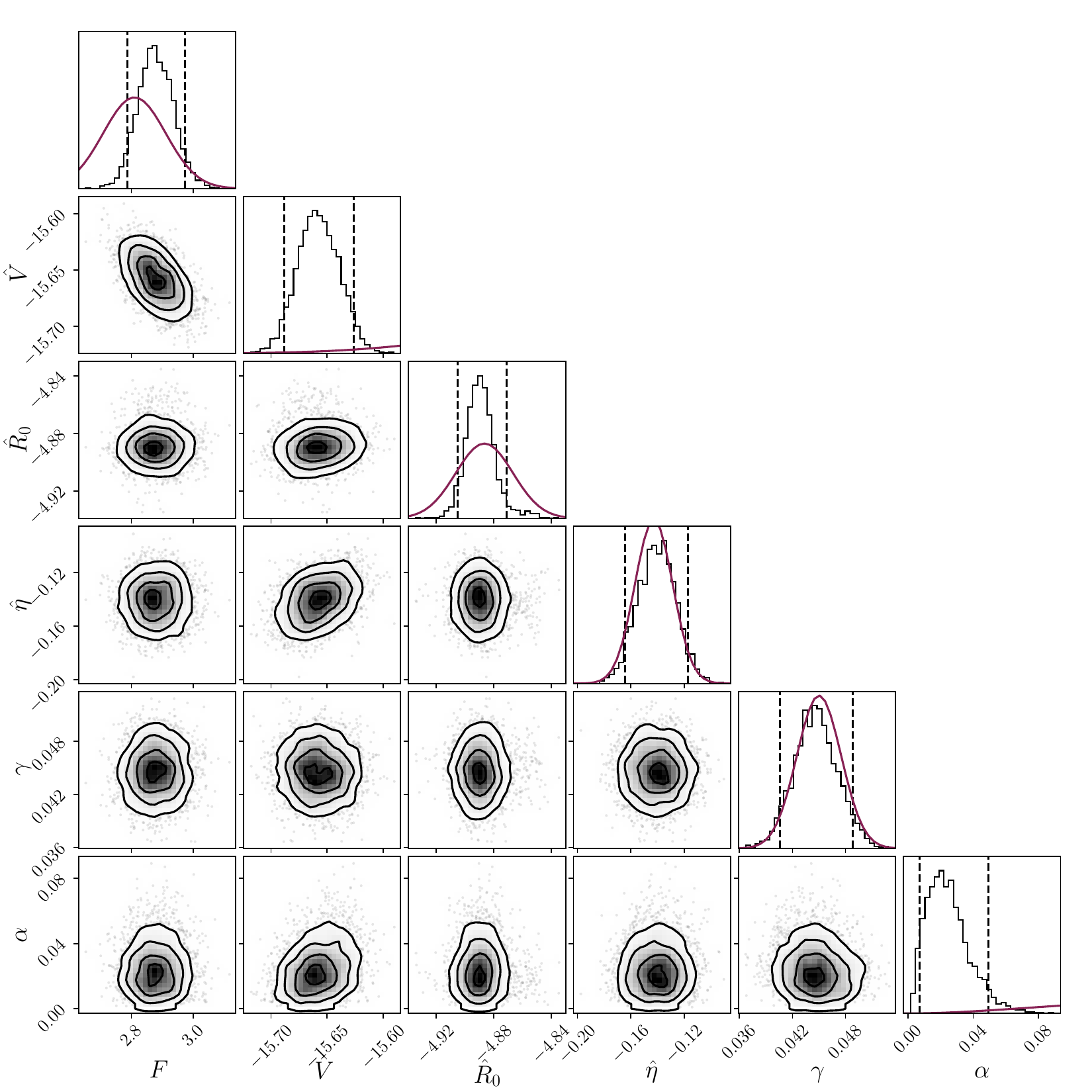}
    \caption{Corner plot showing the correlation between the posterior of the model parameters for case \RNum{1}, where we use glycerol. The histogram and dark-red graphs denote the posterior and prior, respectively. The region in between the black dashed lines defines the 95\% credibility interval.}
    \label{fig:BI_Gly_NLP_corner}
\end{figure*}

\begin{table}
    \centering
    \caption{Uncertainty in the Newtonian squeeze flow model parameters $\boldsymbol{\theta}$ for case \RNum{1}.}
    \begin{tabularx}{\columnwidth}{XXXX}
    \toprule
         &  $\mu$ & $\sigma$ & CV [\%] \\ \midrule
        $F$ [N] & 2.88 & 5.78$\,\times\,10^{-2}$ &  2.01 \\ 
        $V$ [mL] & 1.59$\,\times\,10^{-1}$ & 2.95$\,\times\,10^{-3}$ & 1.86 \\
        $R_0$ [cm] & 7.53$\,\times\,10^{-1}$ & 8.27$\,\times\,10^{-3}$ & 1.10 \\
        $\eta_\mathrm{N}$ [Pa$\cdot$s] & 8.69$\,\times\,10^{-1}$ & 1.24$\,\times\,10^{-2}$ & 1.43 \\ 
        $\gamma$ [N/m] & 4.46$\,\times\,10^{-2}$ & 2.49$\,\times\,10^{-3}$ & 5.59 \\
        $\alpha$ [-] & 2.48$\,\times\,10^{-2}$ & 1.13$\,\times\,10^{-2}$ & 5.36$\,\times\,10^{1}$ \\ \bottomrule
    \end{tabularx}
    \label{tab:musig_BI_squeezeGly}
\end{table}

\subsection{Generalized Newtonian fluid}
We now consider the comparison between the model predication and experimental data for PVP. The results for the experimental cases \RNum{5} to \RNum{8} are visualized in Figures~\ref{fig:UP_PVP_NLP01} and \ref{fig:UP_PVP_TPLLP01}. We observe that cases \RNum{5} and \RNum{6} show a larger initial radius in comparison to cases \RNum{7} and \RNum{8}, which can be explained by the increased fluid volume in cases \RNum{5} and \RNum{6}. Comparing cases \RNum{5} and \RNum{6}, we see that the radius in the latter case increases faster than in the former, which is related to the larger applied force in case \RNum{6}. Similar observations and conclusions are made between cases \RNum{7} and \RNum{8}, where the latter shows the larger increase in radius. The uncertainty in experimental data increases over time as shown in Figure~\ref{fig:UP_PVP_NLP01} and \ref{fig:UP_PVP_TPLLP01}, similar to the observations made for the experimental data of glycerol (see \autoref{fig:UP_Gly_NLP01}).

\subsubsection*{Uncertainty propagation}
The parametric uncertainties of $F$, $V$, $R_0$ and $\gamma$ have been determined through additional experiments and the uncertainty in the rheological parameters $\lambda$, $\eta_0$, $\eta_\infty$ and $n$ have been determined using Bayesian inference applied on the simple shear measurements. These uncertainties are propagated through the squeeze flow model to predict the outward motion of the fluid. We have used 12,240 samples from each distribution of the model parameters to retrieve 12,240 samples for the radius per point in time.   

We start by comparing the Newtonian squeeze flow model to the experimental data obtained for PVP solution, including and excluding Laplace pressure. We choose to implement the viscosity at the zero-shear rate plateau as input for the Newtonian squeeze flow model. We expect to observe a mismatch between the data and model prediction, because of the shear thinning flow behavior at high shear rates (see \autoref{fig:rheodataPVP}).  

\begin{figure}
    \centering
    \includegraphics[width=\columnwidth, trim={0.5cm 1.25cm 1.25cm 0}, clip]{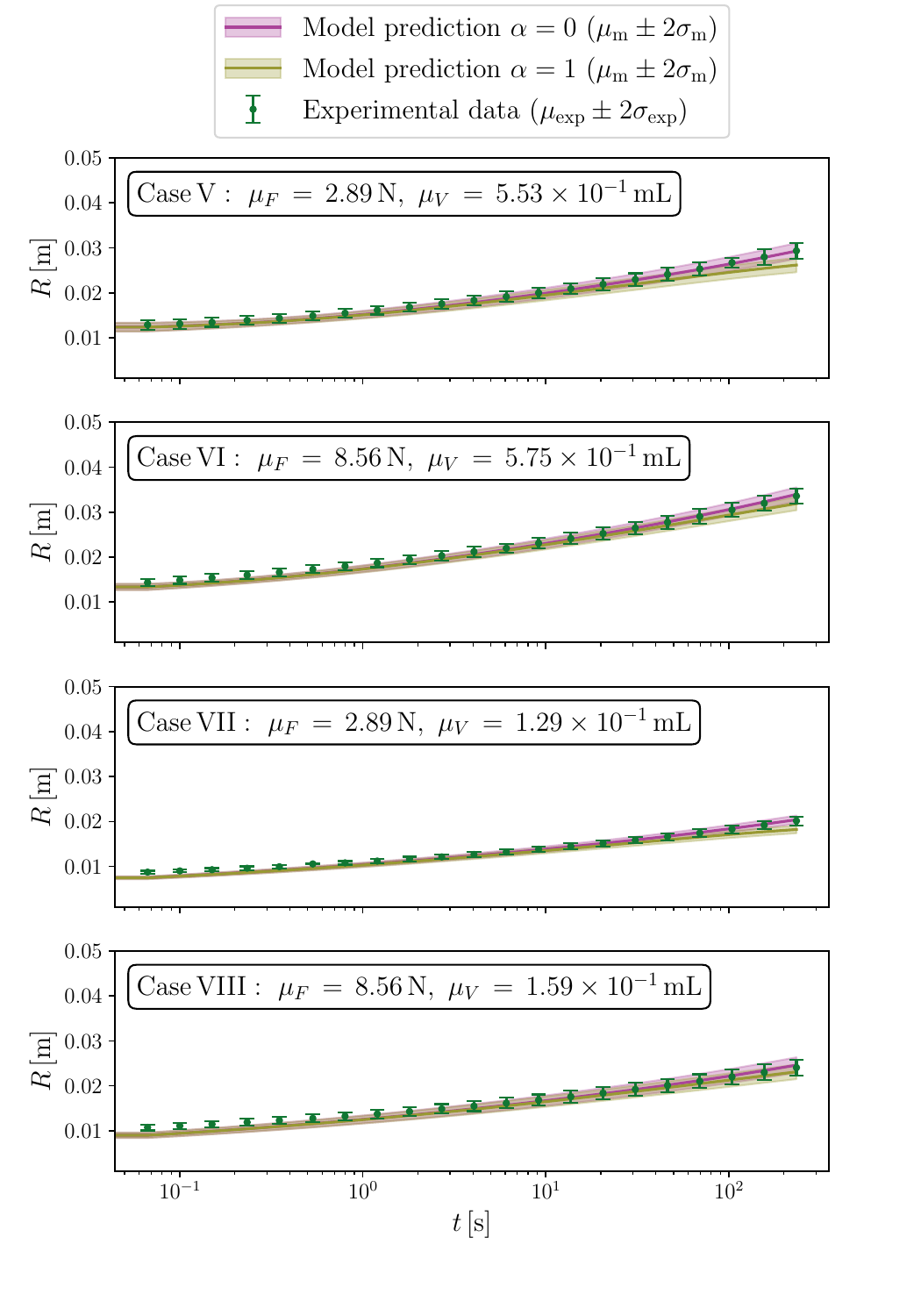}
    \caption{Newtonian model prediction compared to the experimental squeeze flow data of PVP using uncertainty propagation, where the subscript 'm' corresponds to the model and the subscript 'exp' to the experimental data. }
    \label{fig:UP_PVP_NLP01}
\end{figure}

In \autoref{fig:UP_PVP_NLP01}, we show the Newtonian model prediction and experimental data for PVP. There is no significant distinction between the model prediction including and excluding the Laplace pressure. Furthermore, we see that initially the experimental data is underestimated by the model prediction. This can be explained by the constant viscosity at the zero shear rate plateau. Due to the high initial shear rates, the viscosity is lower than the viscosity at the zero-rate plateau, meaning that the evolution rate of the fluid front is higher. We expect that the model prediction improves as we incorporate shear thinning flow behavior in the model through the truncated power law model. 

\begin{figure}
    \centering
    \includegraphics[width=\columnwidth, trim={0 1cm 1cm 0}, clip]{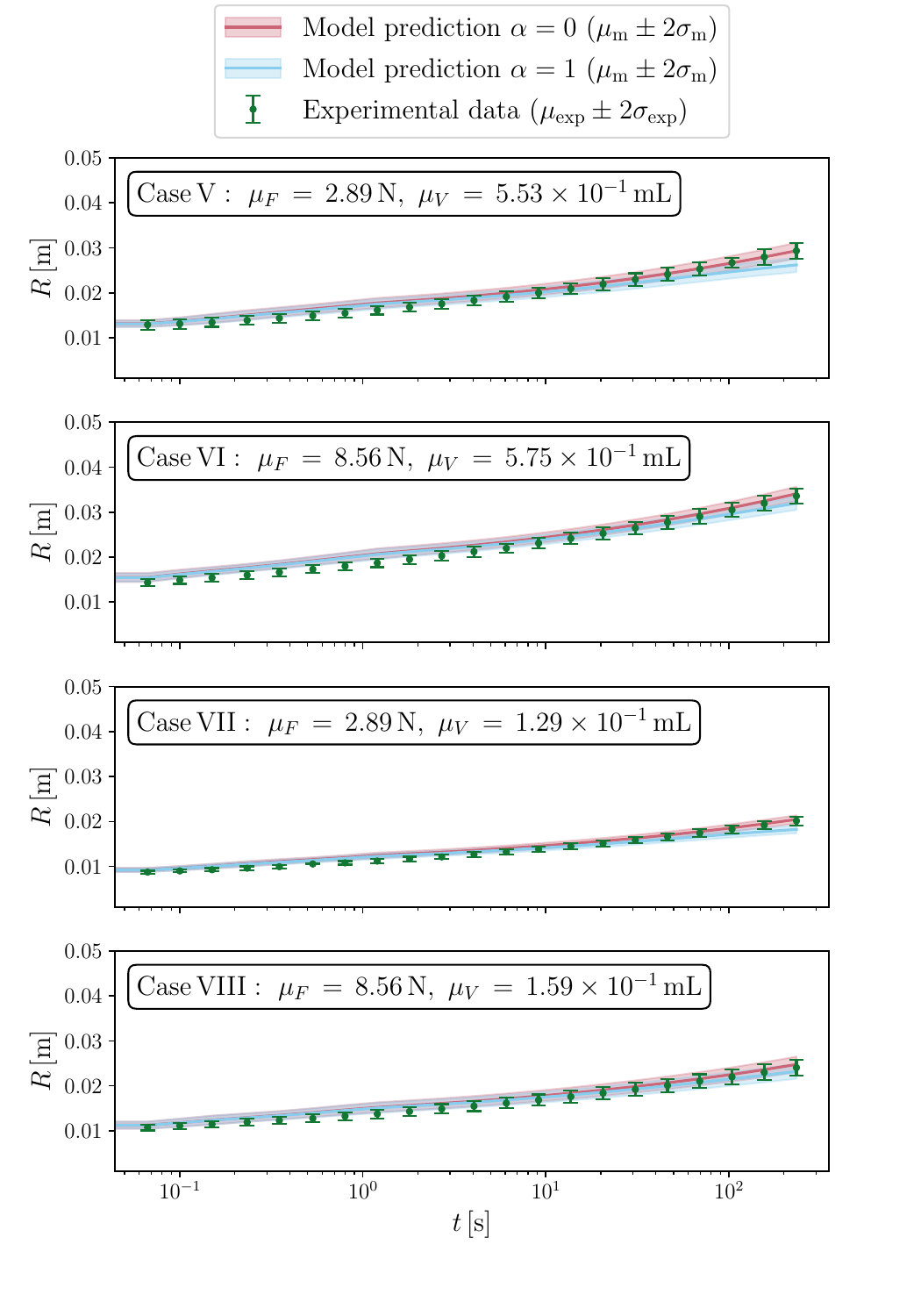}
    \caption{Truncated power law model prediction compared to the experimental squeeze flow data of PVP using uncertainty propagation. The subscript 'm' denotes to the model and the subscript 'exp' to the experimental data}
    \label{fig:UP_PVP_TPLLP01}
\end{figure}

The truncated power law model prediction for $\alpha=0$ and $\alpha=1$ is visualized in \autoref{fig:UP_PVP_TPLLP01}. We observe that for the initial stages the model prediction has improved using the truncated power law model instead of the Newtonian model. As time proceeds, the model prediction describes the experimental data similarly as the Newtonian model prediction, which can be explained by the decrease in shear rate. Around $t=1\,\mathrm{s}$ for cases \RNum{5} to \RNum{8}, the model overestimates the experimental data, independent of the inclusion of the Laplace pressure.

We can use Bayesian inference to update the rheological parameters using the rheological data as prior knowledge. In this way, we can investigate whether the rheological parameters as determined in simple flow are representative for the squeeze flow. Furthermore, we will use this method to investigate the uncertainty in curvature of the interface at the fluid front. We will not apply Bayesian inference on the Newtonian model separately, since it is a special case (\emph{i.e.}, specific parameter setting) of the truncated power law model.

\subsubsection*{Bayesian inference}
We apply the framework of Bayesian inference, using MCMC, on the squeeze flow model with the truncated power law model incorporated. The prior distribution of the model input parameters are similar to the ones used in uncertainty propagation. The rheological parameters, $\eta_0$, $\eta_\infty$, $n$ and $\lambda_{\mathrm{cr}}$, should be non-negative and therefore are assigned a lognormal distribution (\autoref{rmk:logtransform}). The surface tension and applied force have a normal distribution and we use a lognormal distribution like the one used for the rheological parameters for the fluid volume and initial radius. The prior distribution of the curvature is a Beta-distribution, given by equation~\eqref{eq:betadist}, where $a=3$, $b=8$.

We use 27 walkers/chains, 10,000 samples per walker, a burn-in period of $2\times\tau_f$ and $\tau_f$/2 for the thinning to apply Bayesian inference on the squeeze flow model. The autocorrelation time for cases \RNum{5} to \RNum{8} is provided in \autoref{tab:autocor_PVP} per model parameter. The effective sample size is given in \autoref{tab:Neff_PVP} for cases \RNum{5} to \RNum{8}.
\begin{table}
    \centering
    \caption{Autocorrelation times per model parameter and case using 27 walkers and 10,000 samples per walker.}
    \begin{tabularx}{\columnwidth}{XXXXXXXXXX}
    \toprule
         Case &  $F$ & $\hat{V}$ & $\hat{R}_0$ & $\hat{\eta}_0$ & $\hat{\eta}_\infty$ & $n$ & $\lambda_\mathrm{cr}$ & $\gamma$ & $\alpha$ \\ \midrule
        \RNum{5} & 111 & 95 & 111 & 97 & 113 & 120 & 101 & 105 & 111 \\ 
        \RNum{6} & 113 & 98 & 115 & 94 & 101 & 111 & 97 & 104 & 130  \\
        \RNum{7} & 114 & 124 & 135 & 127 & 127 & 132 & 137 & 129 & 170 \\ 
        \RNum{8} & 96 & 95 & 110 & 107 & 115 & 106 & 107 & 115 & 136 \\ \bottomrule
    \end{tabularx}    
    \label{tab:autocor_PVP}
\end{table}

\begin{table}
    \centering
    \caption{Effective sample size $N_\mathrm{final}$ per case based on the highest autocorrelation time per case, for sample size $N=270,000$. }
    \begin{tabularx}{\columnwidth}{XXXXX}
    \toprule
        Case & \RNum{5} & \RNum{6} & \RNum{7} & \RNum{8} \\ \midrule
        $N_\mathrm{final}$ & 4374 & 4023 & 3267 & 3915 \\ \bottomrule
    \end{tabularx}
    \label{tab:Neff_PVP}
\end{table}

\begin{figure}
    \centering
    \includegraphics[width=\columnwidth, trim={0.5cm 1cm 0.5cm 0cm}, clip]{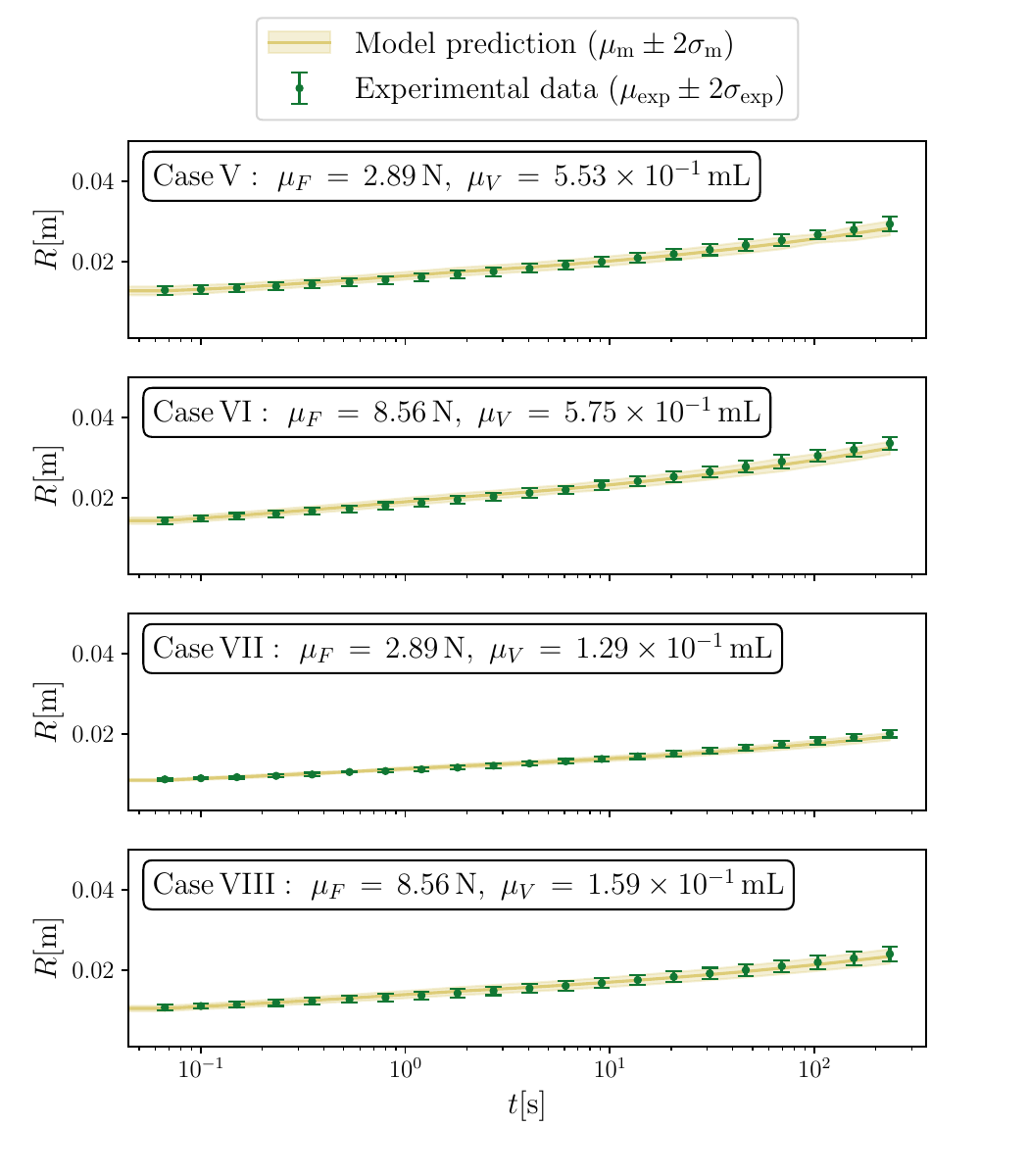}
    \caption{Truncated power law model prediction compared to the experimental squeeze flow data of PVP using Bayesian inference. The subscript 'm' corresponds to the model and the subscript 'exp' to the experimental data}
    \label{fig:BI_PVP_TPLLP01}
\end{figure}

\begin{figure*}
    \centering
    \includegraphics[width=\textwidth]{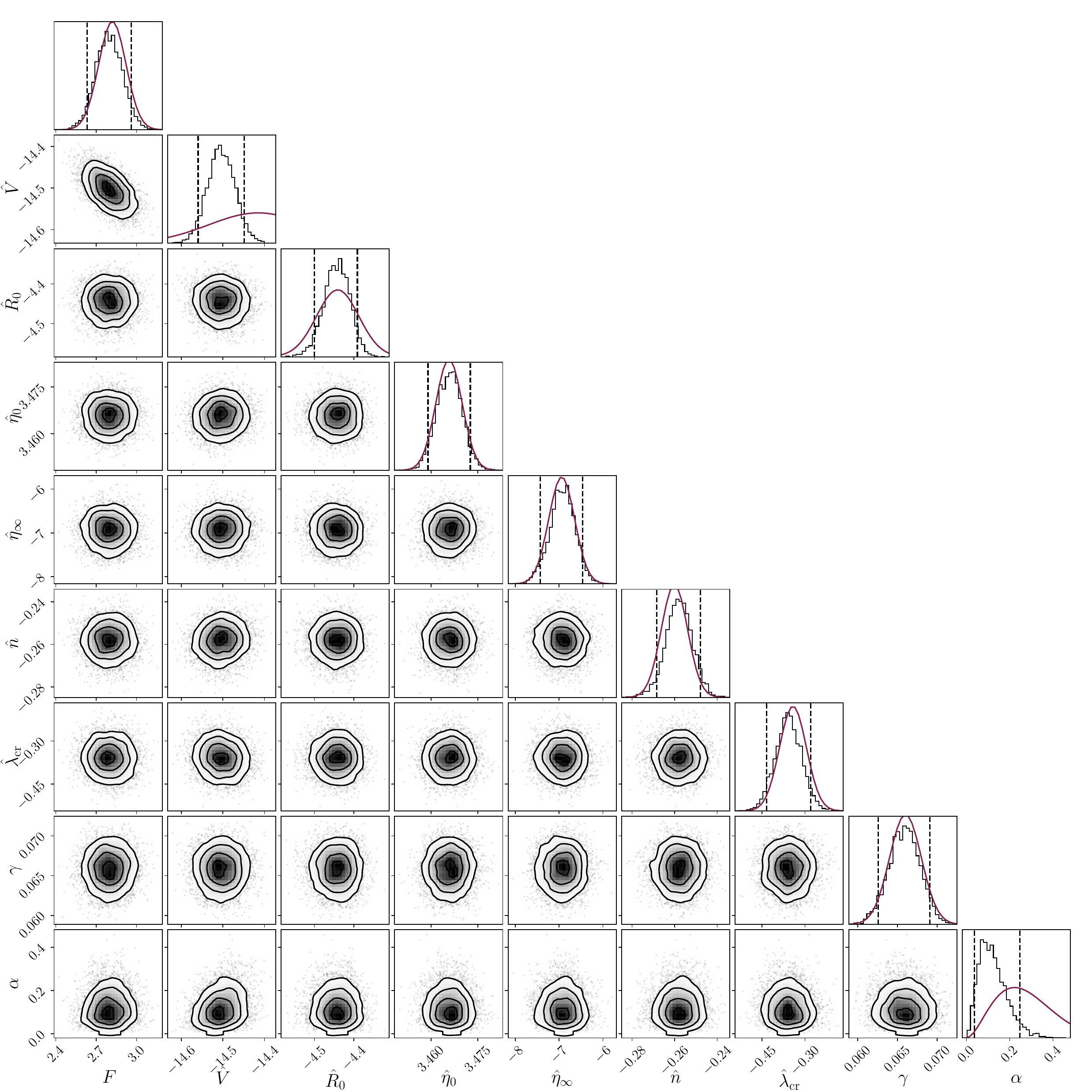}
    \caption{Corner plot showing the correlation between the posterior of every model parameter for case \RNum{5}, where we use PVP solution. The histogram and dark-red graphs denote the posterior and prior, respectively. The region in between the black dashed lines defines the 95\% credibility interval.}
    \label{fig:BI_PVP_TPLLP_corner}
\end{figure*}

In \autoref{fig:BI_PVP_TPLLP01}, the experimental cases \RNum{5} to \RNum{8}  and model predictions using the Bayesian inference framework are given. The mismatch between the experimental data and model prediction around $t=1\,\mathrm{s}$ is significantly decreased in the model prediction using Bayesian inference, meaning that the data available through the likelihood improved the model prediction.
To get a closer look at the posterior distribution for the individual parameters, we analyze the two-dimensional posterior marginals provided in \autoref{fig:BI_PVP_TPLLP_corner}. The posterior distributions are given by the histograms on the far right image of every row, including the prior distribution given in red. We observe that the posterior distribution corresponds to the prior distribution for the applied force $F$ and rheological parameters $\hat{\eta}_0$ and $\hat{\eta}_\infty$ and reasonably corresponds to the rheological parameters $\hat{n}$ and $\hat{\lambda}_\mathrm{cr}$. The posterior distribution of $\gamma$ is similar to the prior distribution and the prior distribution does not correspond to the posterior distribution of $\alpha$. Note that $\alpha$ is shifted toward zero in comparison to the prior distribution, but not as strong as in the glycerol case (\autoref{fig:BI_Gly_NLP_corner}). The posterior of the initial radius corresponds to the prior distribution, where the posterior distribution is more narrow than the prior distribution and the mean remains approximately the same. Using Bayesian inference led to a decrease in uncertainty in the initial radius. A significant mismatch can be observed in the prior and posterior distribution of the fluid volume. The mismatch in prior and posterior distribution of the parameters underlies the improved prediction using Bayesian inference around $t=1\,\mathrm{s}$ in \autoref{fig:BI_PVP_TPLLP01} in comparison to \autoref{fig:UP_PVP_TPLLP01} for cases \RNum{5} to \RNum{8}. 

In analyzing the correlation between the parameters, we investigate the two-dimensional posterior marginals provided in \autoref{fig:BI_PVP_TPLLP_corner}. We observe that nearly every posterior marginal does not show a significant correlation between parameters. The posterior marginal between the applied force $F$ and fluid volume $V$ is elliptic, which could indicate a correlation between the parameters or a difference in level of informativeness in the prior. For the parameters to which we have applied the log-transformation, the posteriors are given by $\hat{\boldsymbol{\theta}}$ in the corner plot. The statistical moments are given in \autoref{tab:musig_BI_squeezePVP} for the actual model parameters $\boldsymbol{\theta}$. The uncertainty in the rheological parameters and surface tension is expected to be similar for cases \RNum{5} to \RNum{8}. In \ref{app:uncertaintytab} we show the uncertainty in the case independent parameters ($\eta_0$, $\eta_\infty$, $n$, $\lambda_\mathrm{cr}$ and $\gamma$) for the remaining cases (\RNum{6} to \RNum{8}), from which it can be concluded that the uncertainty per case is comparable as expected.

\begin{table}
    \centering
    \caption{Uncertainty in the truncated power law squeeze flow model parameters $\boldsymbol{\theta}$ for case \RNum{5}.}
    \begin{tabularx}{\columnwidth}{XXXX}
    \toprule
         &  $\mu$ & $\sigma$ & CV [\%] \\ \midrule
        $F$ [N] & 2.80 & 1.00$\,\times\,10^{-1}$ &  3.60 \\ 
        $V$ [mL] & 5.03$\,\times\,10^{-1}$ & 1.68$\,\times\,10^{-2}$ & 2.99 \\
        $R_0$ [cm] & 1.18 & 3.88$\,\times\,10^{-2}$ & 3.30 \\
        $\eta_0$ [Pa$\cdot$s] & 3.20$\,\times\,10^{1}$ & 1.32$\,\times\,10^{-1}$ & 4.13$\,\times\,10^{-1}$\\
        $\eta_\infty$ [Pa$\cdot$s] & 1.02$\,\times\,10^{-3}$ & 3.04$\,\times\,10^{-4}$ & 2.97$\,\times\,10^{1}$\\
        $n$ [-] & 7.73$\,\times\,10^{-1}$ & 4.87$\,\times\,10^{-3}$ & 6.30$\,\times\,10^{-1}$\\
        $\lambda_\mathrm{cr}$ [s] & 7.00$\,\times\,10^{-1}$ & 3.28$\,\times\,10^{-2}$ & 4.68 \\
        $\gamma$ [N/m] & 6.59$\,\times\,10^{-2}$ & 1.97$\,\times\,10^{-3}$ & 2.99 \\
        $\alpha$ [-] & 1.24$\,\times\,10^{-1}$ & 6.47$\,\times\,10^{-2}$ & 5.22$\,\times\,10^{1}$ \\ \bottomrule
    \end{tabularx}
    \label{tab:musig_BI_squeezePVP}
\end{table}

\subsubsection*{Model-based results}
In this section we analyze model-based results for the PVP solution, \emph{i.e.}, results acquired for non-observable quantities. We analyze two model-based results at $t=1.0\,\times\,10^{-2}$\,s,  $t=1.0$\,s, $t=10$\,s and $t=30$\,s: the local viscosity regimes and the velocity profile. The results are shown for case \RNum{5}. Using the data available in the supplementary information, the results can be obtained for cases \RNum{6} to \RNum{8} as well.

\begin{figure}
    \centering
    \includegraphics[width=\columnwidth]{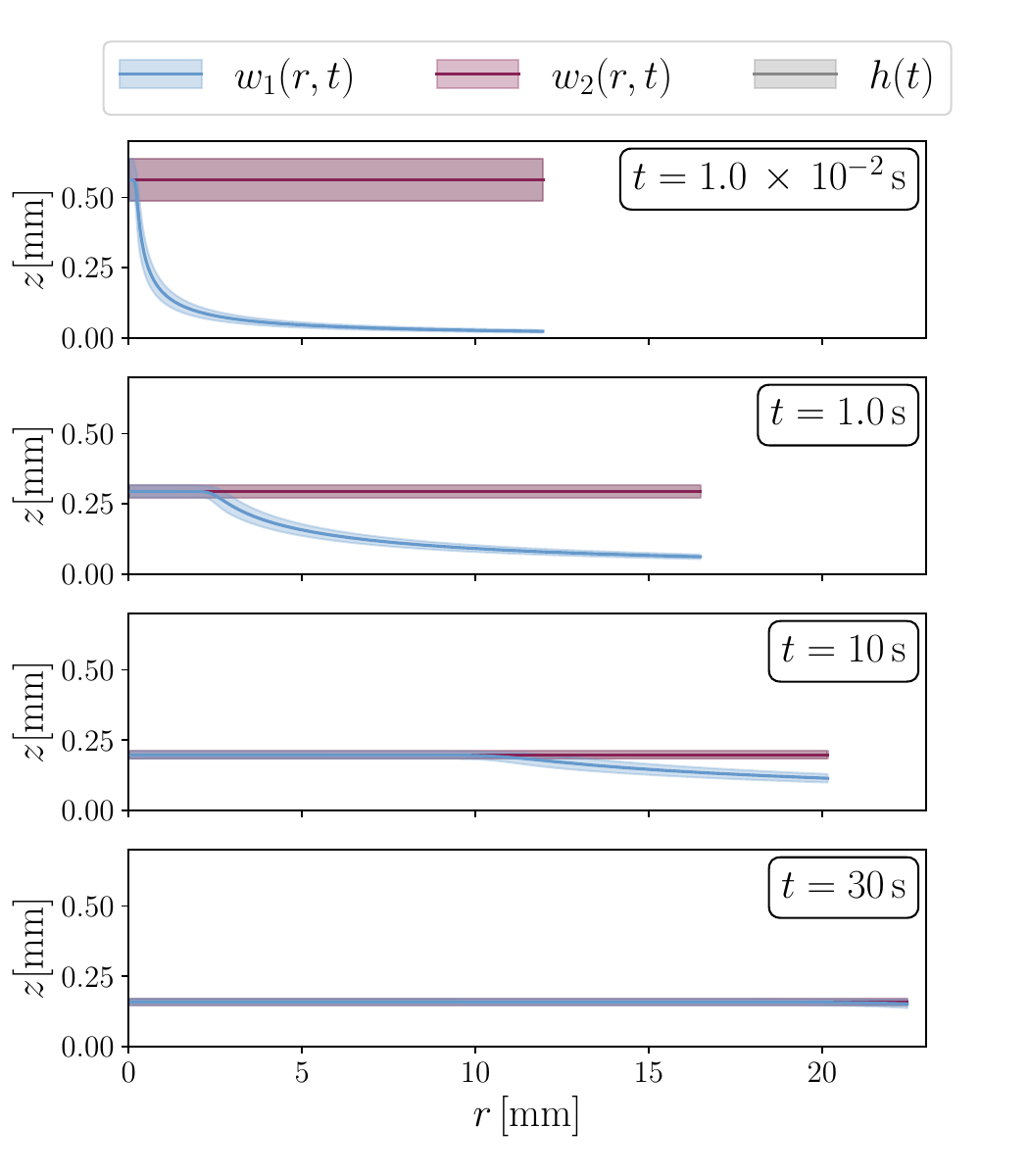}
    \caption{Uncertainty in the local viscosity regimes of case \RNum{5} for various points in time. The error bars cover the $\mu \pm 2 \sigma$ region. Due to the relatively small uncertainty in $r$ in comparison to the uncertainty in $w_1(r,t)$, $w_2(r,t)$ and $h(t)$, we plot the mean of the distribution of $r$.}
    \label{fig:w1w2ht531}
\end{figure}

First, we discuss the local viscosity regimes. In Section~\ref{sec:model}, we have discussed the three regimes in the truncated power law model: Newtonian at zero-shear rate plateau, power law at intermediate shear rates and Newtonian at infinite shear rates. Using the squeeze flow model, we can obtain the interfaces $w_1(t,r)$, $w_2(t,r)$ and $h(t)$. In addition, we can obtain the uncertainty in the position of the interfaces. In \autoref{fig:w1w2ht531}, we show each of the interfaces with the uncertainty. Note that the domain is half of the height of the fluid layer. In this case, we do not reach a high enough shear rate to reach the third regime, where $\eta=\eta_\infty$, therefore, $w_2(r,t)=h(t)$, due to which we cannot distinguish these two regimes in the figure. In time, we observe a decrease in height and increase in radius as expected. By comparing the local viscosity regime for $t=1.0\,\times\,10^{-2}$\,s to $t=30$\,s we observe that the area covered by the Newtonian regime, where $\eta=\eta_0$, increases, whereas the power law regime decreases. This shows that the shear thinning behavior decreases through time. The uncertainty in $w_1(r,t)$ is not constant in the radial domain. In time, the level of uncertainty in $h(t)$ decreases.

\begin{figure}
    \centering
    \includegraphics[width=\columnwidth]{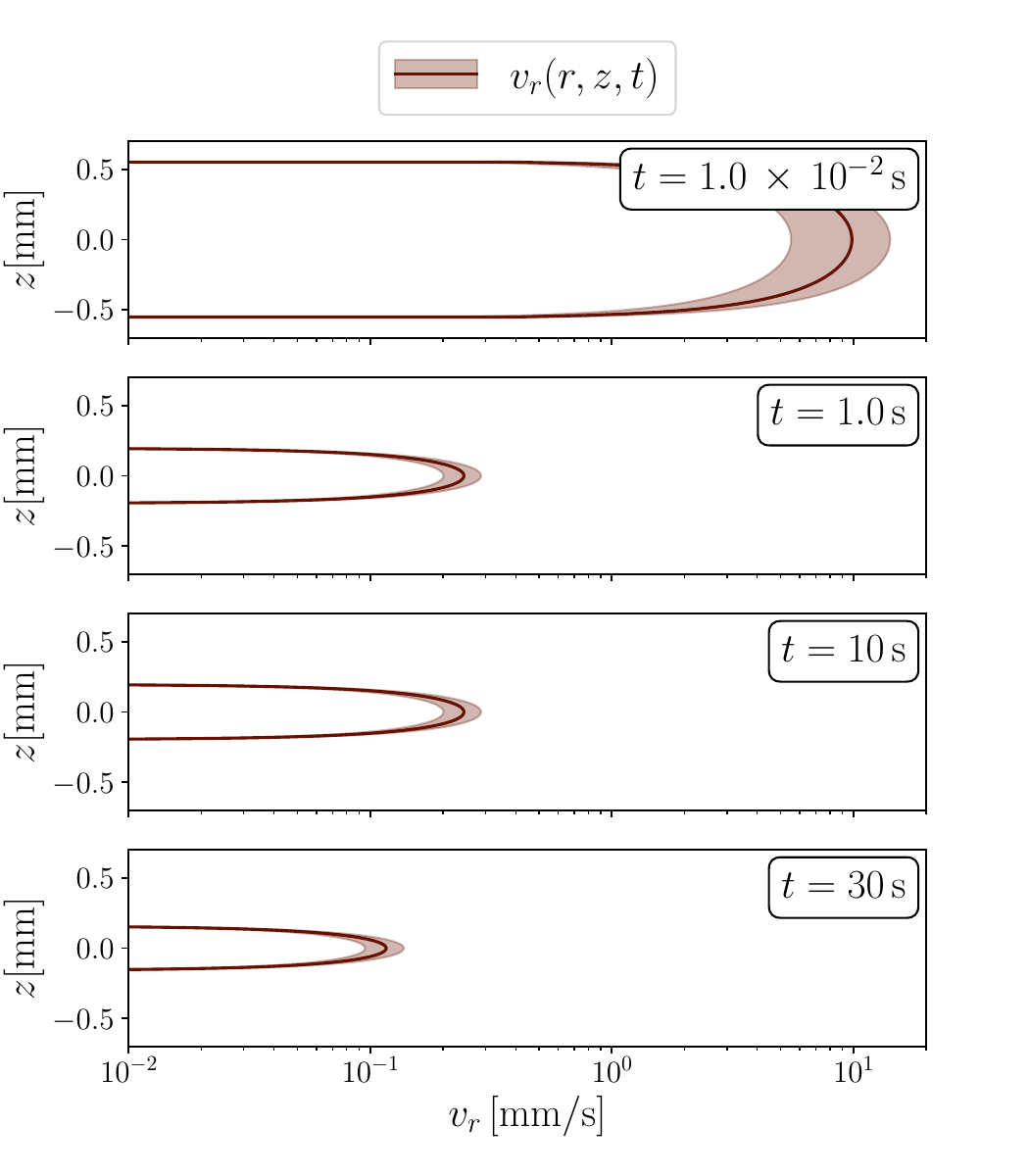}
    \caption{Uncertainty in the velocity profile of case \RNum{5} for various points in time. The error bars cover the $\mu \pm 2 \sigma$ region. Due to the relatively small uncertainty in $r$ in comparison to the uncertainty in $v_r(r,z,t)$, we plot the mean of the distribution of $R$. Furthermore, the uncertainty in $h$ is relatively small in comparison to the uncertainty in $v_r(r,z,t)$ due to which we plot the mean of the distribution of $h$.}
    \label{fig:velproft531}
\end{figure}

In \autoref{fig:velproft531} we show the velocity profile at $r=R$ for $t=1.0\,\times\,10^{-2}$\,s,  $t=1.0$\,s, $t=10$\,s and $t=30$\,s. Note that we show the full height of the fluid layer. We observe a rapid decrease in velocity between $t=1.0\,\times\,10^{-2}$\,s and $t=1.0$\,s, after which the velocity slightly decreases. Furthermore, the height decreases over time in the squeeze flow and we show that the height decreases the fastest in the initial stages. Near the walls, the uncertainty is equal to zero because we assume a no-slip condition. Toward the middle of the velocity profile, the level of uncertainty increases. We expect the profile to change with time, because the shear thinning behavior is only present during the initial stages. By comparing the model prediction of the velocity for the four different points in time, we do not directly observe the changes in the form of the velocity profile, which we attribute to the logarithmic scale and/or the decrease in height. The level of uncertainty decreases in time, especially from $t=1.0\,\times\,10^{-2}$\,s to $t=1.0$\,s.

\section{Conclusions and recommendations}
\label{sec:conclusion}

We have used UQ to make probabilistic predictions for generalized Newtonian fluids in squeeze flow. Two methods to quantify the uncertainty in model response have been applied: uncertainty propagation and Bayesian inference. For uncertainty propagation, the parametric uncertainty is propagated through the squeeze flow model to obtain the uncertainty in model response. For Bayesian inference, the experimental squeeze flow data is used to update the model parameters, treated as probabilistic in this framework, and thereby obtain an improved model prediction.
To enable the costly inference step using MCMC sampling, we have developed a semi-analytical model, assuming a three-regime truncated power law for the viscosity.
The UQ framework allows us to separate measurement noise and model bias, and thus make decisions as to which model is more suitable for the
given experimental observations.
We have seen that for a PVP solution during the initial stages of the squeeze flow,  shear thinning behavior is significant, due to which the truncated power law prediction is superior to the Newtonian prediction. Furthermore, the Laplace pressure boundary condition at the fluid front should be included for improved predictions during the later stages. Moreover, we can  quantify model-based predictions for quantities that are not directly observable, such as the velocity profile and the local viscosity regimes. With the obtained results, we demonstrate the capabilities of the UQ framework in the field of non-Newtonian fluid mechanics.

Because the concept of uncertainty quantification of non-Newtonian fluid mechanics is in its infancy, we recognize that many paths for future work are possible.
For example, an optimization of the MCMC sampling method could lead to a significant reduction in computational time and improve the accuracy of the uncertainty in model parameters and model predictions. Furthermore, efficient parallel sampling methods would be essential in applying UQ using more advanced numerical models. 
 With respect to the obtained squeeze flow result, advanced rheological models and fluids can be incorporated, such as viscoelastic and viscoplastic models and fluids. Other model extensions that could be incorporated are, \emph{e.g.}, wall-slip and thermal effects. This raises the question whether a more complex model is justified given the experimental data, which we intend to adress in future work using Bayesian model selection.

\section*{Acknowledgements}
\label{sec:acknowledgements}

We acknowledge Michiel van Gorp from the Eindhoven University of Technology for the assistance in developing the tailored experimental setup. In addition, we thank Coen van der Gracht from the Eindhoven University of Technology for conducting the pendant drop measurements.

\appendix
\section{Lubrication theorem}
\label{sec:APPscaling}

In this appendix we elaborate the dimensional reduction of the model equations based on the lubrication theory, which applies to our problem on account of the fluid layer height ($H$) being much smaller than the radius ($R$). We first consider the mass and momentum balances, followed by a discussion of the shear rate.

\subsection*{Balance laws}\label{app:scalemommass}
We consider the balance laws, equations~  \eqref{ch3_mass_bal_1} and \eqref{ch3_mom_bal_1}, expressed in cylinder coordinates ($r$, $\theta$, $z$), where the $\theta$-direction is neglected based on the axisymmetry assumption:
\begin{subequations} \label{APPmod_setmassmom}
    \begin{align}
    \frac{1}{r}\frac{\partial}{\partial r}(r v_r)+\frac{\partial v_z}{\partial z} & = 0 \label{APPmod_massbal}\\
        \frac{\partial p}{\partial r} & = \frac{1}{r}\frac{\partial}{\partial r}\left(r \tau_{rr}\right)+\frac{\partial \tau_{rz}}{\partial z} \label{APPmod_mombalr}\\
        \frac{\partial p}{\partial z} & = \frac{1}{r}\frac{\partial}{\partial r}\left(r \tau_{zr}\right)+\frac{\partial \tau_{zz}}{\partial z}\label{APPmod_mombalz}
    \end{align}
\end{subequations}
In these expressions, $\tau_{rr}$, $\tau_{rz}$, $\tau_{zr}$ and $\tau_{zz}$ are the $rr$-component, $rz$-component, $zr$-component and the $zz$-component of the extra stress tensor, respectively. In \autoref{APPmod_setmassmom}, $p$, $v_r$ and $v_z$ are the pressure, $r$-component of the velocity and $z$-component of the velocity. The parameters in \autoref{APPmod_setmassmom} can be scaled as
\begin{align}\label{ch4_scale_par}
\begin{split}
    & r^* = \frac{r}{R_0}, \quad z^* = \frac{z}{H_0}, \quad p^* = \frac{p}{F}\pi R_0^2, \\
    & v_r^*=\frac{v_r}{V_r}, \quad v_z^* = \frac{v_z}{V_z}, \quad t^* = \frac{t}{t_c}, \\
    & \tau_{rr}^* = \frac{\tau_{rr}}{
    \tau_{rr,c}}, \quad \tau_{rz}^* = \frac{\tau_{rz}}{
    \tau_{rz,c}}, \quad \tau_{zr}^* = \frac{\tau_{zr}}{
    \tau_{zr,c}}, \quad \tau_{zz}^* = \frac{\tau_{zz}}{
    \tau_{zz,c}},
\end{split}
\end{align}
where the subscript $0$ refers to the initial state of the system, 
\begin{align}
 V_r &=  \frac{F H_0^2}{\pi R_0^3} , & V_z &= \frac{H_0}{R_0} V_r =  \frac{F H_0^3}{\pi R_0^4} ,
\end{align}
and the characteristic values for the components of the extra stress are 
\begin{equation}\label{eq:scaletau}
    \tau_{rr,c} = \frac{V_r}{R_0}, \quad \tau_{rz,c} = \frac{V_r}{H_0}, \quad \tau_{zr,c} = \frac{V_z}{R_0}, \quad \tau_{zz,c} = \frac{V_z}{H_0}.
\end{equation}
Upon substitution of these scaling relations, the balance laws \eqref{APPmod_setmassmom} read
\begin{subequations}
\begin{align}
&\frac{1}{r^*}\frac{\partial}{\partial r^*}(r^* v_r^* )+ \frac{\partial v_z^*}{\partial z^*} = 0,\\
&\frac{\partial p^*}{\partial r^*} = \frac{H_0^2}{R_0^2} \frac{1}{r^*}\frac{\partial }{\partial r^*} \left(r  \tau_{rr}^*\right) + \frac{\partial^2 \tau_{rz}^*}{\partial z^*{}^2},\\
&\frac{\partial p^*}{\partial z^*} = \frac{H_0^4}{R_0^4}\frac{1}{r^*}\frac{\partial}{\partial r^*}\left(r^*\tau_{zr}^*\right) + \frac{H_0^2}{R_0^2}\frac{\partial^2 \tau_{zz}^*}{\partial z^*{}^2}.
\end{align}%
\end{subequations}
From the lubrication assumption, $H_0 \ll R_0$, it then follows that the pressure is constant in the $z$-direction and that
\begin{subequations}
\begin{align}
    \frac{1}{r}\frac{\partial}{\partial r}\left(r v_r\right) - \frac{\partial v_z}{\partial z} & = 0,  \\
    \frac{\partial}{\partial z}\tau_{rz} & = \frac{\partial p}{\partial r}.
\end{align}
\end{subequations}

\subsection*{Shear rate}\label{app:scaleshear}
The shear rate is defined as
\begin{equation}
    \dot{\gamma} = \sqrt{2\boldsymbol{D}:\boldsymbol{D}},
    \label{eq:appshearrate}
\end{equation}
where $\boldsymbol{D}$ is the rate-of-deformation tensor, which, upon consideration of the axisymmetry condition, can be expressed in the cylindrical coordinate system as
\begin{subequations}
    \begin{align}
        D_{rr} & = \frac{\partial v_r}{\partial r}, \\
        D_{zz} & = \frac{\partial v_z}{\partial z}, \\ 
        D_{zr} & = D_{rz} = \frac{1}{2} \left[ 
\frac{\partial v_z}{\partial r} + \frac{\partial v_r}{\partial z} \right].
    \end{align}    
\end{subequations}
Substitution of the scaling relations defined above then yields
\begin{subequations}
\begin{align}
    D_{rr} & = \frac{V_r}{R_0} \frac{\partial v_r^*}{\partial r^*}, \\
    D_{zz} & = \frac{V_r}{R_0} \frac{\partial v_z^*}{\partial z^*}, \\
    D_{zr} & = D_{rz} = \frac{1}{2} \left[ \frac{V_r H_0}{R_0^2} \frac{\partial v_z^*}{\partial r^*} + \frac{V_r}{H_0} \frac{\partial v_r^*}{\partial z^*} \right].
\end{align} 
\end{subequations}
Using the lubrication assumption, $H_0 \ll R_0$, the absolute shear rate \eqref{eq:appshearrate} can finally be rewritten as
\begin{equation}
    \dot{\gamma} = \sqrt{\left( 2 D_{rz} \right)^2} = \left| \frac{\partial v_r}{\partial z} \right|.
\end{equation}

\section{Non-linear squeeze flow solver}
\label{sec:APPsolver}

We employ a nonlinear time integrator to attain the solution of the squeeze flow model in the case that the truncated power law is considered (see Section~\ref{subsec:TPL}). The essential idea of this solver is to integrate the balance of mass \eqref{eq:TPLstrong} to obtain the total mass flux \eqref{eq:TPLtotalflux} as
\begin{equation}
    Q = -r \dot{h},
    \label{eq:integratedmassbalance}
\end{equation}
where $r$ and $\dot{h}$ are the radial coordinate and rate of domain semi-height, respectively. From equations \eqref{eq:TPLfluxes} and \eqref{eq:TPLws} it follows that the total flux can be expressed as a function of the pressure gradient, $p'=\frac{\partial p}{\partial r}$, as
\begin{equation}
    Q(p') = -\kappa(p') p', 
    \label{eq:conductivity}
\end{equation}
where the proportionality constant, $\kappa$, is a function of the local pressure gradient through the region intervals \eqref{eq:TPLws}. Combining equations \eqref{eq:integratedmassbalance} and \eqref{eq:conductivity} then yields
\begin{equation}
    p' = \frac{r \dot{h}}{\kappa(p')}, 
    \label{eq:fixedpoint}
\end{equation}
which can be integrated over the domain to obtain
\begin{equation}
  p(r)  = \Delta p - \dot{h} \int_r^R \frac{\hat{r}}{\kappa(p'(\hat{r}))} \,{\rm d}\hat{r},
  \label{eq:pressureintegral}
\end{equation}
where $\hat{r}$ is the integration variable. The semi-height rate, $\dot{h}$, is eliminated from this expression using 
\begin{equation}
    \int_{0}^R p(r) r\,{\rm d}r = \frac{R^2}{2}  \Delta p  -  \dot{h} \int_{0}^R   \left[ \int_r^R \frac{\hat{r}}{\kappa(p'(\hat{r}))} \,{\rm d}\hat{r} \right] r\,{\rm d}r = \frac{F}{2 \pi}
\end{equation}
to obtain
\begin{equation}
       \dot{h}  =  \left( \frac{R^2}{2}  \Delta p - \frac{F}{2 \pi} \right) \left[ \int_{0}^R   \left[ \int_r^R \frac{\hat{r}}{\kappa(p'(\hat{r}))} \,{\rm d}\hat{r} \right] r\,{\rm d}r \right]^{-1}.
\end{equation}
Noting that this expression essentially yields the semi-height rate as a function of the pressure-gradient field, equation \eqref{eq:fixedpoint} in essence has the form of a fixed point iteration, which can be solved for the pressure gradient field, and, through \eqref{eq:pressureintegral} for the pressure field.

To evaluate the integral in equation \eqref{eq:pressureintegral} we employ the midpoint rule on a discretization of the domain $[0,R]$ in $n_{\rm int}$ segments. We use backward Euler time integration, which implies that the radius $R$ in the above derivation also varies during the fixed point iterations.

The initial time step is determined based on the Newtonian model in the high shear rate regime ($\eta=\eta_\infty$) as
\begin{equation}
    \Delta t = s\cdot\frac{3 \pi \mu_\infty R_0^4}{8 F h_0^2},
\end{equation}
where $s$ is a scaling parameter (typically $s=0.001$). We use an adaptive time step because in the initial stages of the squeeze flow simulation the time step is required to be much smaller than in the later stages due to high shear rates. The new time step $\Delta t_{i+1}$ is dependent on the Picard iterations and the current time step $\Delta t_i$ as 
\begin{equation}
    \Delta t_{i+1} = \Delta t_i \frac{n_{\mathrm{target}}}{n_{\mathrm{iter}}},
\end{equation}
where $n_{\mathrm{target}}$ is the number of desired Picard iterations (typically $n_{\mathrm{target}}=20$) and $n_{\mathrm{iter}}$ is the number of Picard iterations used in the previous time step. 

\subsection*{Convergence study}
We here investigate the mesh convergence and time step convergence of our non-linear solver. In \autoref{tab:inputconvergence} the truncated power law parameter settings used for the convergence simulations are listed.

\begin{figure}
     \centering
     \subfloat[]{\includegraphics[width=0.5\columnwidth]{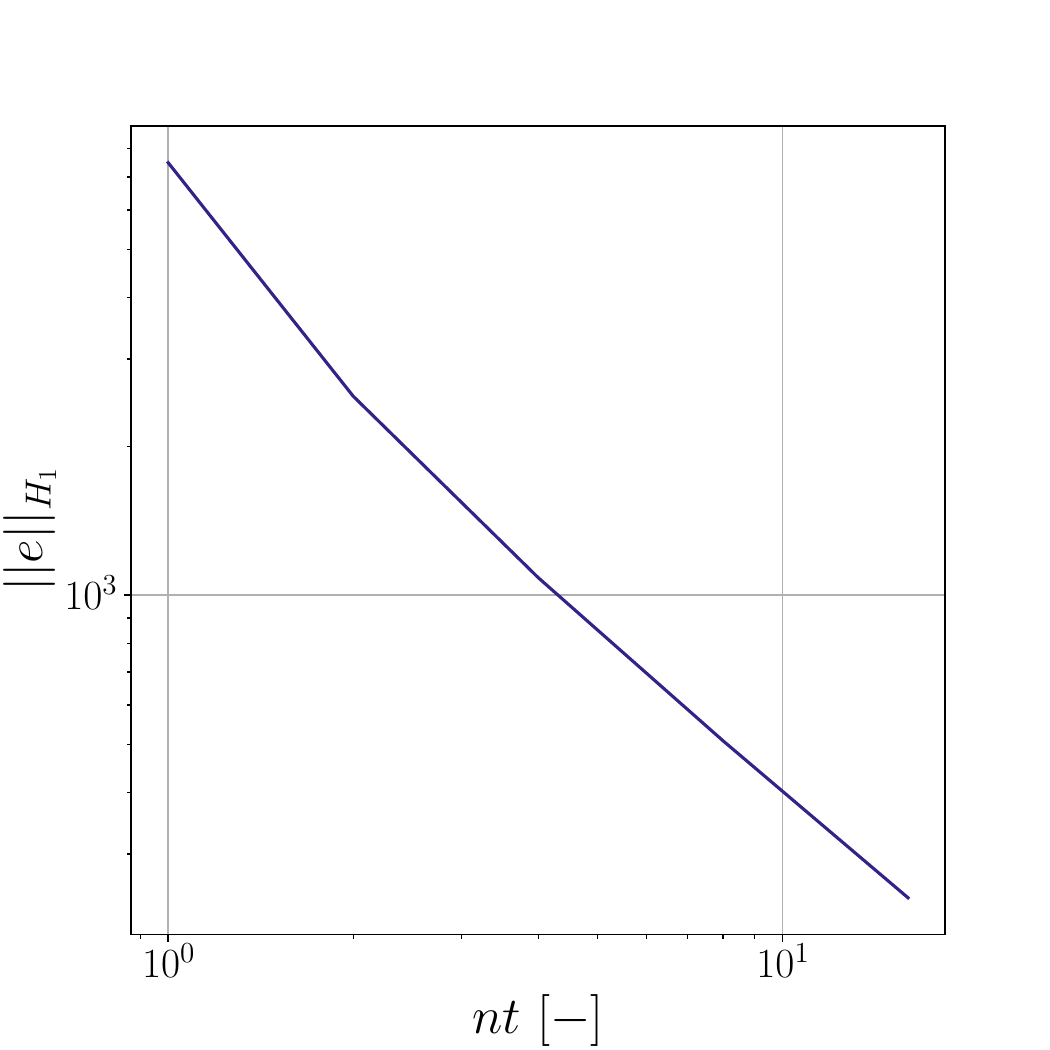}\label{fig:dtH1}}
     \hfill
     \subfloat[]{\includegraphics[width=0.5\columnwidth]{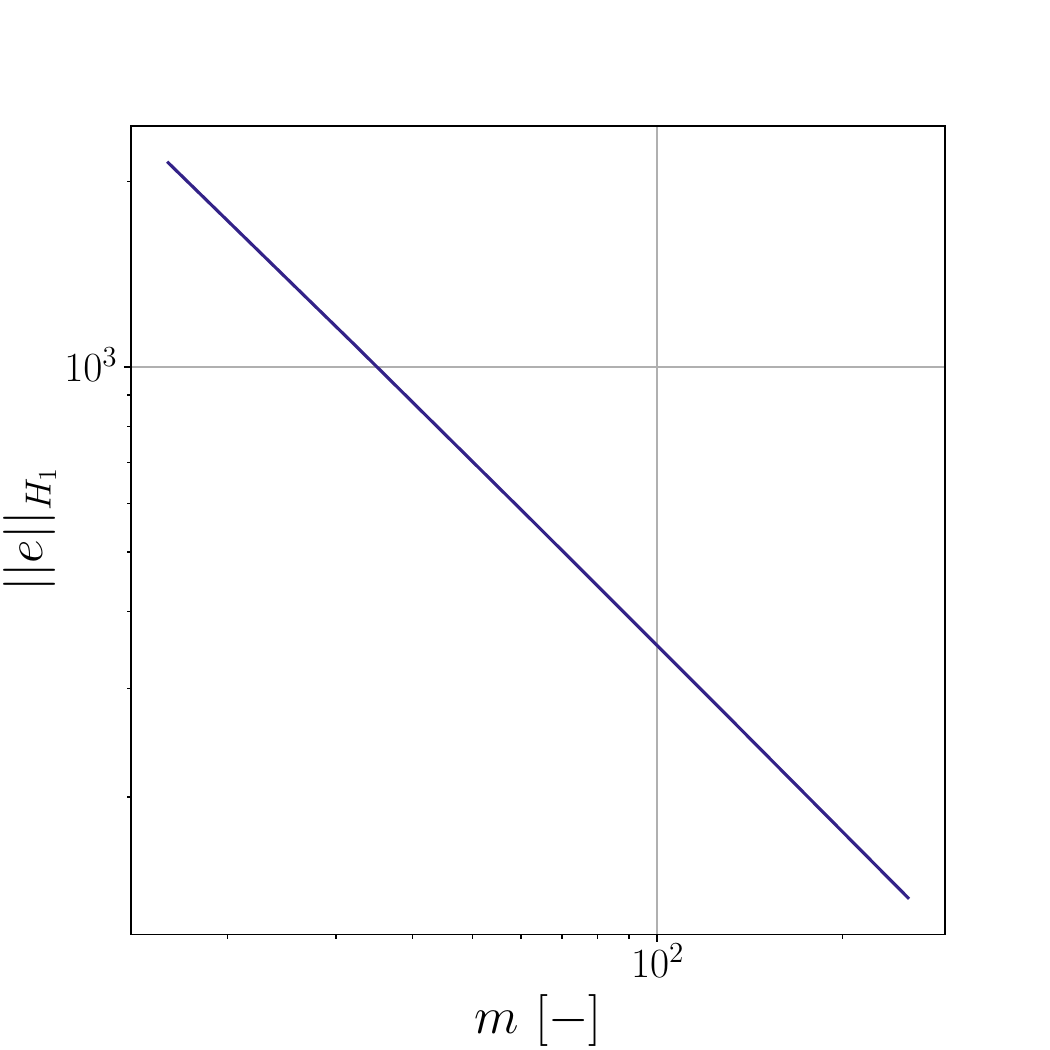}\label{fig:mH1}}
     \caption{Convergence study based on the pressure error $||e||_{H_1}$; a) time step convergence, b) mesh convergence.}
     \label{fig:convergencestudy}
\end{figure}

\begin{table}
    \centering
    \caption{Model input parameters for a truncated power law model used in the convergence study.}
    \begin{tabularx}{\columnwidth}{XXXXXXX}
    \toprule
         $F [\mathrm{N}]$ & $V [\mathrm{mL}]$ & $R_0 [\mathrm{cm}]$ & $\eta_0 \mathrm{Pa} \cdot \mathrm{s}]$ & $\eta_\infty [\mathrm{Pa} \cdot \mathrm{s}]$ & $n [-]$ & $\lambda_\mathrm{cr} [s]$\\ \midrule
         10 & 0.63 & 2.0 & 1.0 & 0.10 & 0.63 & 0.89 \\ \bottomrule
    \end{tabularx}
    \label{tab:inputconvergence}
\end{table}
For the convergence study, we define the error for the pressure field as
\begin{equation}
    e = p - p^h,
\end{equation}
where $p$ is the analytical solution to which the approximate pressure solution $p^h$ is compared. Because no analytical solution can be obtained for the pressure, $p$ is defined using a very fine mesh (16,384 elements) and a high number of time steps (1,024). In \autoref{tab:dtconvergence}, five simulation inputs are given with an increase in number of time steps. In \autoref{tab:mconvergence}, five simulation inputs are provided with an increase in number of elements. 
\begin{table}
   \centering
           \captionsetup[subtable]{position = below}
          \captionsetup[table]{position=top}
           \caption{Convergence study inputs for the number of time steps $n_t$ and number of elements $m$.}
           \begin{subtable}{0.5\linewidth}
               \centering
               \begin{tabularx}{0.9\columnwidth}{XXX}
                   \toprule
                   Case & $m \ [-]$ & $n_t \ [-]$ \\ \midrule
                   1 & 16,384 & 1 \\
                   2 & 16,384 & 2 \\
                   3 & 16,384 & 4 \\
                   4 & 16,384 & 8 \\ 
                   5 & 16,384 & 16 \\ \bottomrule 
               \end{tabularx}
               \caption{}
               \label{tab:dtconvergence}
           \end{subtable}
           \begin{subtable}{0.5\linewidth}
               \centering
               \begin{tabularx}{0.9\columnwidth}{XXX}
                   \toprule
                   Case & $m \ [-]$ & $n_t \ [-]$ \\ \midrule
                   6 & 16 & 1,024 \\
                   7 & 32 & 1,024 \\
                   8 & 64 & 1,024 \\
                   9 & 128 & 1,024 \\ 
                   10 & 256 & 1,024 \\ \bottomrule
               \end{tabularx}
               \caption{}
                 \label{tab:mconvergence}
           \end{subtable}
\end{table}
We investigate the H$_1$-error of the pressure field for the number of time steps and the number of elements to analyze the convergence in the pressure $p$ and the gradient of the pressure $\frac{\partial p}{\partial r}$, given by
\begin{equation}
    ||e||_{H_1} = \sqrt{\int_0^R e^2 + \left( 
\frac{\partial e}{\partial r} \right)^2 \, \mathrm{d}r}.
\end{equation}
In \autoref{fig:dtH1} the time step convergence is presented. The rate of convergence deviates for less than two time steps, which is caused by the time step being too coarse, resulting in pre-asymptotic behavior. In \autoref{fig:mH1} the mesh convergence is visualized. We observe asymptotic convergence behavior for both the mesh size and time step size. A detailed convergence study is beyond the scope of this work.

\section{Uncertainty in the squeeze flow model parameters using Bayesian inference}
\label{app:uncertaintytab}

In this section we evaluate the uncertainty obtained for the squeeze flow model parameters that are case independent using Bayesian inference. We expect that the uncertainty in rheological parameters and surface tension for cases \RNum{1} to \RNum{4} is similar, even though we applied Bayesian inference on the separate cases. The same hypothesis holds for cases \RNum{5} to \RNum{8}. 
We show the uncertainty in the Newtonian squeeze flow model parameters in \autoref{tab:app_musig_BI_squeezeGly} and the truncated power law squeeze flow model parameters in \autoref{tab:app_musig_BI_squeezePVP}. We observe a similar uncertainty in each of the model parameters. 
\begin{table}
    \centering
    \caption{Uncertainty in case independent Newtonian squeeze flow model parameters $\boldsymbol{\theta}$ for the cases \RNum{2} to \RNum{4}.}
    \begin{tabularx}{\columnwidth}{lllll}
    \toprule
        Case &  &  $\mu$ & $\sigma$ & CV [\%] \\ \midrule
        
        \multirow{2}{*}{ \RNum{2}} & \multicolumn{1}{l}{$\eta_\mathrm{N}$ [Pa$\cdot$s]} & \multicolumn{1}{l}{8.70$\,\times\,10^{-1}$}  & \multicolumn{1}{l}{1.24$\,\times\,10^{-2}$} & \multicolumn{1}{l}{1.43} \\
        & \multicolumn{1}{l}{$\gamma$ [N/m]} & \multicolumn{1}{l}{4.46$\,\times\,10^{-2}$}  & \multicolumn{1}{l}{2.45$\,\times\,10^{-3}$} & \multicolumn{1}{l}{5.51} \\ \midrule
            
        \multirow{2}{*}{ \RNum{3}} & \multicolumn{1}{l}{$\eta_\mathrm{N}$ [Pa$\cdot$s]} & \multicolumn{1}{l}{8.69$\,\times\,10^{-1}$}  & \multicolumn{1}{l}{1.24$\,\times\,10^{-2}$} & \multicolumn{1}{l}{1.43} \\
        & \multicolumn{1}{l}{$\gamma$ [N/m]} & \multicolumn{1}{l}{4.47$\,\times\,10^{-2}$}  & \multicolumn{1}{l}{2.53$\,\times\,10^{-3}$} & \multicolumn{1}{l}{5.66} \\ \midrule
        
        \multirow{2}{*}{ \RNum{4}} & \multicolumn{1}{l}{$\eta_\mathrm{N}$ [Pa$\cdot$s]} & \multicolumn{1}{l}{8.83$\,\times\,10^{-1}$}  & \multicolumn{1}{l}{1.24$\,\times\,10^{-2}$} & \multicolumn{1}{l}{1.40} \\
        & \multicolumn{1}{l}{$\gamma$ [N/m]} & \multicolumn{1}{l}{4.48$\,\times\,10^{-2}$}  & \multicolumn{1}{l}{2.41$\,\times\,10^{-3}$} & \multicolumn{1}{l}{5.37} \\ \bottomrule
    \end{tabularx}
    \label{tab:app_musig_BI_squeezeGly}
\end{table}
\begin{table}
    \centering
    \caption{Uncertainty in case independent truncated power law squeeze flow model parameters $\boldsymbol{\theta}$ for the cases \RNum{6} to \RNum{8}.}
    \begin{tabularx}{\columnwidth}{lllll}
    \toprule
        Case &  &  $\mu$ & $\sigma$ & CV [\%] \\ \midrule
        \multirow{5}{*}{ \RNum{6}} & \multicolumn{1}{l}{$\eta_0$ [Pa$\cdot$s]} & \multicolumn{1}{l}{3.20$\,\times\,10^{1}$}  & \multicolumn{1}{l}{1.32$\,\times\,10^{-1}$} & \multicolumn{1}{l}{4.13$\,\times\,10^{1}$} \\
       
       & \multicolumn{1}{l}{$\eta_\infty$ [Pa$\cdot$s]} & \multicolumn{1}{l}{9.89$\,\times\,10^{-4}$}  & \multicolumn{1}{l}{2.90$\,\times\,10^{-4}$} & \multicolumn{1}{l}{2.94$\,\times\,10^{1}$} \\
       
       & \multicolumn{1}{l}{$n$ [-]} & \multicolumn{1}{l}{7.74$\,\times\,10^{-1}$}  & \multicolumn{1}{l}{4.76$\,\times\,10^{-3}$} & \multicolumn{1}{l}{6.15$\,\times\,10^{1}$} \\
       
       & \multicolumn{1}{l}{$\lambda_\mathrm{cr}$ [s]} & \multicolumn{1}{l}{6.98$\,\times\,10^{-1}$}  & \multicolumn{1}{l}{3.32$\,\times\,10^{-2}$} & \multicolumn{1}{l}{4.75} \\
       
        & \multicolumn{1}{l}{$\gamma$ [N/m]} & \multicolumn{1}{l}{6.59$\,\times\,10^{-2}$}  & \multicolumn{1}{l}{2.02$\,\times\,10^{-3}$} & \multicolumn{1}{l}{3.07} \\ \midrule

        \multirow{5}{*}{ \RNum{7}} & \multicolumn{1}{l}{$\eta_0$ [Pa$\cdot$s]} & \multicolumn{1}{l}{3.20$\,\times\,10^{1}$}  & \multicolumn{1}{l}{1.32$\,\times\,10^{-1}$} & \multicolumn{1}{l}{4.13$\,\times\,10^{1}$} \\
       
       & \multicolumn{1}{l}{$\eta_\infty$ [Pa$\cdot$s]} & \multicolumn{1}{l}{9.93$\,\times\,10^{-4}$}  & \multicolumn{1}{l}{3.07$\,\times\,10^{-4}$} & \multicolumn{1}{l}{3.09$\,\times\,10^{1}$} \\
       
       & \multicolumn{1}{l}{$n$ [-]} & \multicolumn{1}{l}{7.74$\,\times\,10^{-1}$}  & \multicolumn{1}{l}{4.71$\,\times\,10^{-3}$} & \multicolumn{1}{l}{6.08$\,\times\,10^{1}$} \\
       
       & \multicolumn{1}{l}{$\lambda_\mathrm{cr}$ [s]} & \multicolumn{1}{l}{6.95$\,\times\,10^{-1}$}  & \multicolumn{1}{l}{3.23$\,\times\,10^{-2}$} & \multicolumn{1}{l}{4.65} \\
       
        & \multicolumn{1}{l}{$\gamma$ [N/m]} & \multicolumn{1}{l}{6.59$\,\times\,10^{-2}$}  & \multicolumn{1}{l}{1.97$\,\times\,10^{-3}$} & \multicolumn{1}{l}{2.99} \\ \midrule

        \multirow{5}{*}{ \RNum{8}} & \multicolumn{1}{l}{$\eta_0$ [Pa$\cdot$s]} & \multicolumn{1}{l}{3.20$\,\times\,10^{1}$}  & \multicolumn{1}{l}{1.33$\,\times\,10^{-1}$} & \multicolumn{1}{l}{4.18$\,\times\,10^{1}$} \\
       
       & \multicolumn{1}{l}{$\eta_\infty$ [Pa$\cdot$s]} & \multicolumn{1}{l}{1.00$\,\times\,10^{-3}$}  & \multicolumn{1}{l}{3.03$\,\times\,10^{-4}$} & \multicolumn{1}{l}{3.01$\,\times\,10^{1}$} \\
       
       & \multicolumn{1}{l}{$n$ [-]} & \multicolumn{1}{l}{7.72$\,\times\,10^{-1}$}  & \multicolumn{1}{l}{4.78$\,\times\,10^{-3}$} & \multicolumn{1}{l}{6.19$\,\times\,10^{1}$} \\
       
       & \multicolumn{1}{l}{$\lambda_\mathrm{cr}$ [s]} & \multicolumn{1}{l}{7.08$\,\times\,10^{-1}$}  & \multicolumn{1}{l}{3.37$\,\times\,10^{-2}$} & \multicolumn{1}{l}{4.76} \\
       
        & \multicolumn{1}{l}{$\gamma$ [N/m]} & \multicolumn{1}{l}{6.59$\,\times\,10^{-2}$}  & \multicolumn{1}{l}{1.98$\,\times\,10^{-3}$} & \multicolumn{1}{l}{3.01} \\ \midrule
    \end{tabularx}
    \label{tab:app_musig_BI_squeezePVP}
\end{table}

\bibliographystyle{elsarticle-num-names} 
\bibliography{sections/references.bib}

\end{document}